\documentclass[preprint,12pt]{elsarticle}

\usepackage{amssymb}
\usepackage{amsmath,upgreek}
\usepackage{multicol,caption}
\usepackage{float,placeins}
\usepackage[T1]{fontenc}
\usepackage[colorlinks]{hyperref}
\usepackage{lipsum}
\usepackage{longtable}
\usepackage{graphicx}
\usepackage{subfig}
\usepackage{lineno}
\usepackage{hyperref}
\usepackage{geometry}
\usepackage{pdflscape}
\usepackage[normalem]{ulem} 
\usepackage{comment}
\usepackage{multirow}
\usepackage{float}

\journal{New Astronomy}

\def\sax{SAX~J1748.9-2021}
\def\hete{HETE~J1900.1-2455}
\def\aql{Aql~X-1}

\usepackage[nolist,printonlyused]{acronym}
\begin{acronym}[TDMA]
 \acro{LMXB}{Low mass X-ray binary}
 \acrodefplural{LMXB}{Low mass X-ray binaries}
 \acro{AMXP}{accreting millisecond X-ray pulsar}
 \acrodefplural{AMXP}{accreting millisecond X-ray pulsars}
 \acro{PCA}{proportional counter array}
 \acro{ASM}{all sky monitor}
 \acro{RXTE}{the Rossi X-ray timing explorer}
 \acro{MAXI}{monitor of all sky X-ray image}
 \acro{Swift/XRT}{the Swift gamma-ray burst mission/X-ray telescope}
 \acro{Swift}{the Swift gamma-ray burst mission}
 \acro{XSPEC}{an X-ray spectral fitting package}
 \acro{PCU}{proportional counter unit}
 \acro{ISS}{international space station}
 \acro{MSP}{millisecond pulsar}
 \acro{NS}{neutron star}
 \acro{BH}{black hole}
 \acro{FFT}{Fast Fourier Transform}
 \acro{ML}{maximum likelihood}
\end{acronym}

\newcommand\aap{A\&A}                
\newcommand\aaps{A\&AS}              
\newcommand\apj{ApJ}                 
\newcommand\apjl{ApJ}                
\newcommand\apjs{ApJS}               
\newcommand\memsai{Mem. Soc. Astron. Italiana} 
\newcommand\mnras{MNRAS}             
\newcommand\nat{Nature}              
\newcommand\pasa{Publ. Astron. Soc. Australia}  

\begin{document}

\begin{frontmatter}



\title{An Elaborate Search for Coherent Pulsations from Intermittent--AMXPs}
\author[a,b]{Mustafa Turan Sa\u{g}lam}
\author[c,d]{Can G\"ung\"or\corref{mycorrespondingauthor}}
\cortext[mycorrespondingauthor]{Corresponding author}
\ead{gungor.can@istanbul.edu.tr}
\author[a]{Tu\u{g}\c{c}e Kocab{\i}y{\i}k}

\address[a]{\.{I}stanbul University, Institute of Graduate Studies in Sciences, Department of Astronomy and Space Sciences, Beyaz{\i}t, 34119, \.{I}stanbul, Turkey}
\address[b]{Erciyes University, Science Faculty, Department of Astronomy and Space Sciences, Melikgazi, 38030, Kayseri, Turkey }
\address[c]{\.{I}stanbul University, Science Faculty, Department of Astronomy and Space Sciences, Beyaz{\i}t, 34119, \.{I}stanbul, Turkey}
\address[d]{\.{I}stanbul University Observatory Research and Application Center, Beyaz{\i}t, 34119, \.{I}stanbul, Turkey}

\begin{abstract}

We present a detailed systematic pulse search for three Intermittent-Accreting Millisecond X-ray Pulsars (Intermittent-AMXPs), \hete, \sax~\& \aql, via Z$_1^2$ and \ac{ML} 
techniques by using 16 years data of Rossi X-ray Timing Explorer/Proportional Counter Array (RXTE/PCA) in the energy range of 3.0 – 13.0~keV. We first performed a pulse scan 
using the Z$_1^2$ technique in millisecond sensitivities for every 25~s time interval with 1~s shifts to cover all data set around the detected frequencies given in the 
literature. We tracked the Z$_1^2$ power over time and flagged the time intervals exceeding defined threshold levels for each source as \textit{pulse candidates}. The detected 
pulse list throughout our scan has new discoveries while covering the pulsed regions presented in the literature. For a deeper search, using the pulses obtained from the Z$_1^2$ 
method as a probability density function as an input parameter, we re-scanned the time intervals centered on the detected pulse via ML. The detected pulse-on duration via ML is 
slightly longer than the one via Z$_1^2$ method. This phenomenon allows us to argue for the existence of the smooth transition between pulse-on and pulse-off stages. For \sax, we 
also obtained orbital period by using the systematic pulse arrival phase patterns throughput of ML to be 8.76 hours.

\end{abstract}


\end{frontmatter}


\section{Introduction}
\acp{LMXB} are binary systems consisting of two components, which are a compact object, a neutron star (NS) 
or a black hole and a low-mass main sequence star, the \textit{donor} star, which is the source of the 
transferred mass via the Roche-lobe overflow. 
In systems where the compact object is a NS, the transferred material from the donor star to the NS creates 
an accretion disk around the compact object due to the conservation of angular momentum and the geometric 
properties of the system. The material flowing to the inner edge of the accretion disk channelizes through 
the NS poles following magnetic field lines.
The observed X-ray luminosity originates from the gravitational potential energy of the accreted material 
on the poles (for a common review on accreting binaries see \citep{Chaty2022}).
Yet, 339 \acp{LMXB} have been discovered \citep{Fortin2024}, and most of the \acp{LMXB} don't show 
pulsations in their X-ray light curve. Three different scenarios have been proposed in the literature to 
explain the absence of pulsation; \textit{(i)} The NS magnetic field is not strong enough to carry the 
matter from the inner disk to the polar caps \citep{1982Natur.300..728A, 1985Natur.316..239A}. 
\textit{(ii)} The gravitational lensing effect might diminish the impact of the pulses 
\citep{1988ApJ...327..712M}. \textit{(iii)} The X-ray pulses may disappear via the Compton scattering from 
the electron cloud around the NS \citep{1987ApJ...317L..33B, 1988AdSpR...8b.455K}.

A special subgroup of \acp{LMXB}, \acp{AMXP}, exhibits pulses in their X-ray light curve which are discovered at frequencies higher than $\sim$100 Hz during their outburst \citep{1982Natur.300..728A}. 
The material flowing through the magnetic field lines towards the magnetic pole caps of the NS results in the formation of areas called hot spots.
19 members overall 23 known \acp{AMXP}, the X-ray pulsations are persistent while the rest shows occasional pulses, so called Intermittent-\acp{AMXP} \citep{2013MmSAI..84..117B}. The intermittency of the pulsations from these peculiar systems
make them unique labs to study pulse phenomena from \acp{LMXB} since the physical reason of why all \acp{LMXB} do not show pulsation is still an open question.
In the context of this paper, three intermittent-\acp{AMXP}, \aql, \hete~and \sax~which have public \ac{RXTE} observations are studied.


\textbf{\textit{\aql}} is a unique source, due to its pulsation history in which pulses are detected at 
550.25 Hz only for 150s duration \citep{Casella2008} overall 16 years of observation from \ac{RXTE} and 
some more from newer X-ray missions.
The detected pulse frequency is consistent with the burst oscillations 
observed from the system \citep{koyama1981discovery}.
Despite lots of searches by independent groups, no 
pulse signature except the 150~s one has been found. Recently, \citet{2021PASA...38...11B}, reported 
possible pulses in 13 different regions in total, by performing detailed pulse scans after orbital 
correction to the photon arrival times.

\begin{figure}[h]
  \centering
    \includegraphics[width=3.90in, height=2.76in]{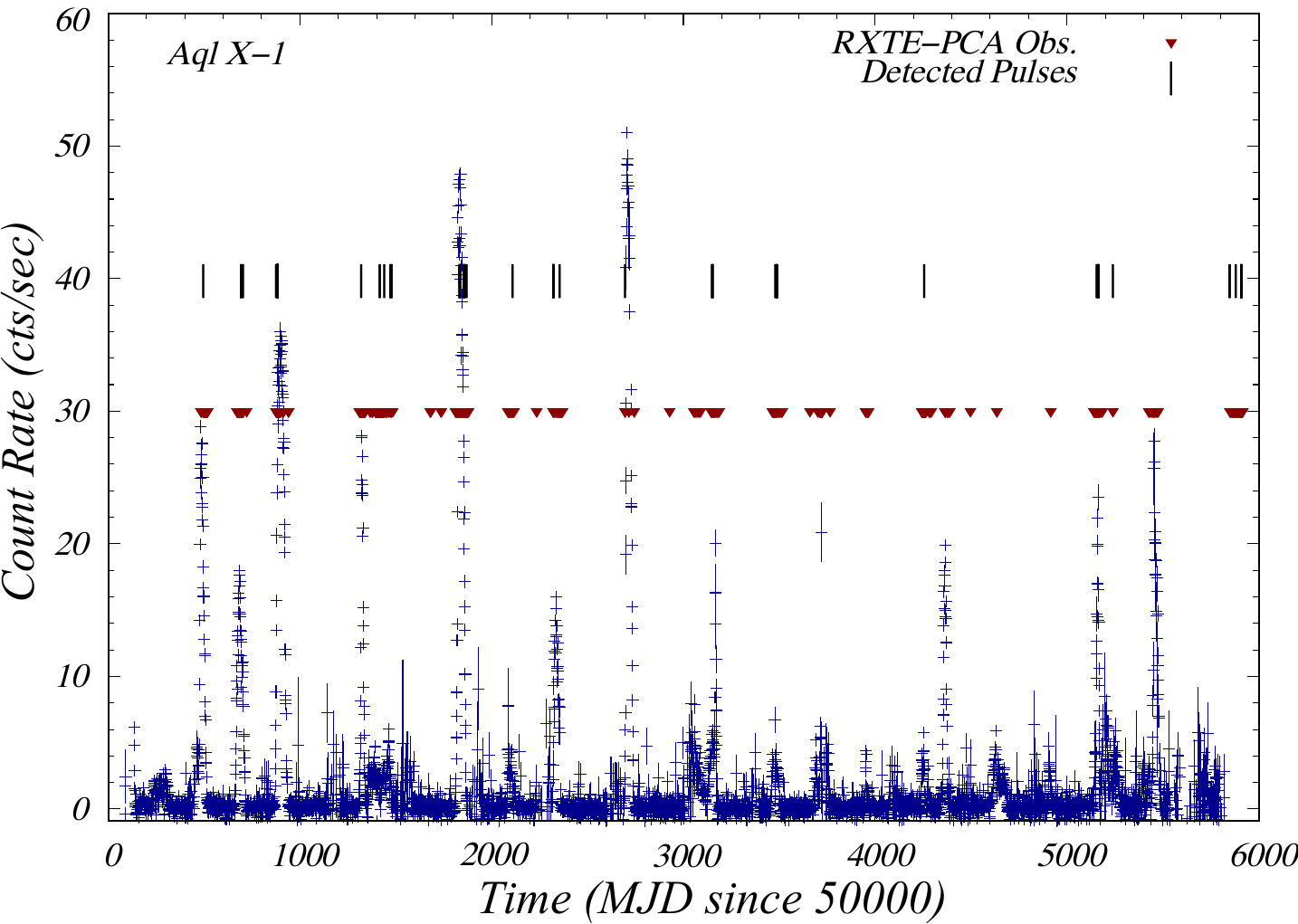}
  \caption{The long term light curve of \aql~via ASM daily data with the times of \ac{RXTE}/\ac{PCA} (upside down brown triangles) as well as the detected pulse candidates (vertical black lines).}
  \label{fig:aql-lc}
\end{figure}

The system has one of the longest orbital period, $\sim$19 hours, of \acp{AMXP} within an 
inclined orbit of $36^\circ$ -- $47^\circ$ \citep{10.1093/mnrasl/slw172}. \aql~shows cyclic outbursts in its X-ray 
light curve almost every year, so the source is also classified as a soft X-ray transient (SXT). 
The distance of the source was reported as 4.4–5.9 kpc via peak flux values during the outburst by \citet{jonker2004distances} via \ac{RXTE} data. In \autoref{fig:aql-lc}, we show the long-term light curve of \aql~from the \ac{ASM} with the time of \ac{RXTE} observations used in this study (brown downwards triangles). 

\begin{figure}[h]
  \centering

\includegraphics[width=3.90in, height=2.76in]{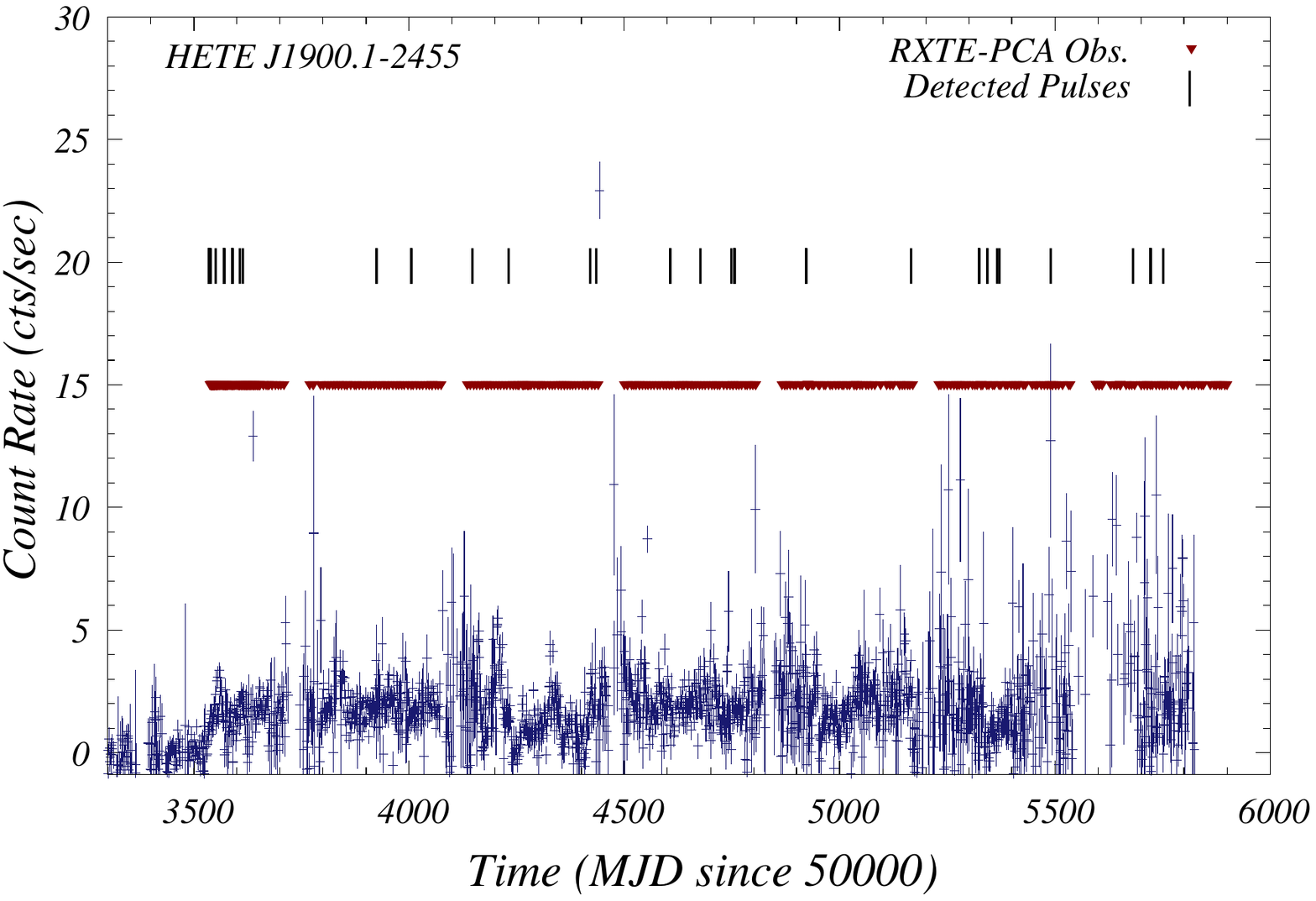}
  \caption{Same as \autoref{fig:aql-lc} but for  \hete.}
  \label{fig:HETE-lc}
\end{figure}

\textbf{\textit{\hete}} was discovered in 2005 by the High Energy Transient Explorer 2 (HETE-2) during a 
very bright X-ray outburst \citep{2005ATel..516....1V}. In the follow-up observations with \ac{RXTE}, 
pulses of the source were detected at 377.3 Hz which led to the classification of the source as an 
\ac{AMXP} \citep{2006ApJ...638..963K}. The observations also revealed that the companion star to the 
neutron star is most likely a brown dwarf with a solar mass of \(0.016-0.07\,M\) and the orbital period of 
the binary system is 83 minutes \citep{2006ApJ...638..963K}.
The source showed pulses for the first 25 days of the 2005 outburst, but then the pulsations suddenly 
disappeared. During the next 2.5 years, sporadic pulsations were observed during outbursts, but no stable 
pulses were detected \citep{GallowayThermo2008}. In \autoref{fig:HETE-lc}, we show the long-term light curve of \hete~from \ac{ASM} data, along with the specific times when \ac{RXTE} 
observations were carried out for this research.

\textbf{\textit{\sax }} is a neutron star X-ray binary, located in the globular cluster NGC 6440. The 
source was discovered in 1998 by BeppoSax during the X-ray activity scan around the Galactic 
center \citep{1999A&A...345..100I}. The distance of the source was determined to 
be \(8.5 \pm 0.4\) kpc \citep{1994A&AS..108..653O}. Pulsations from the source have not been 
detected since the 2005 outburst. 
The frequency of the detected pulse is $\sim$442.3 Hz in a single 
observation from an \ac{RXTE} data \citep{2007ApJ...669L..29G}.
Though, the source is classified as an intermittent-\ac{AMXP}. 

The first estimate of the spin frequency and orbital parameters of \sax~was reported by \citet{AltamiranoSax2008} in the 2001 outburst. \citet{2009ApJ...690.1856P}, using the same data set, but applying a phase-matched timing technique, tried to determine some of the parameters of the source. According to \citet{AltamiranoSax2008}, the companion star could be a main sequence (or slightly evolved) star with a mass ranging from \(0.85\) to \(1.1 \, M_\odot\). In fact, in \citet{2007ApJ...669L..29G} the transient pulsations of this source were analyzed and it was found that the pulsations at \(442.36\) Hz occurred temporarily during the outbursts in 2001 and 2005 and their relation with thermonuclear explosions was discussed.

\begin{figure}[h]
  \centering

    \includegraphics[width=3.90in, height=2.76in]{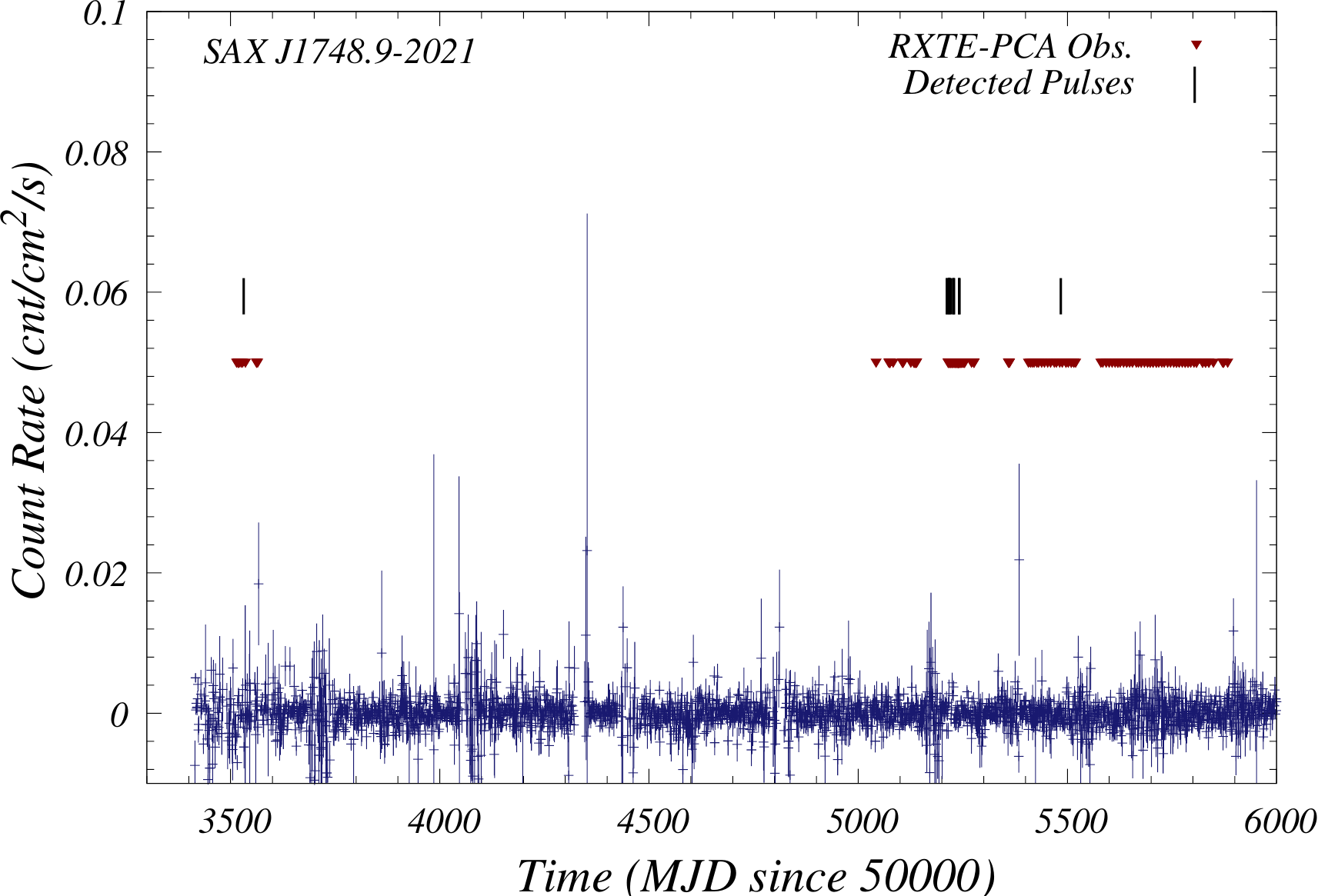}

  \caption{Same as \autoref{fig:aql-lc} but for \sax. The ASM light curve is from the globular cluster NGC 6440.}
  \label{fig:sax-lc}
\end{figure}

Furthermore, \citet{2016MNRAS.459.1340S} reported that new pulses were detected during an outburst in 2015 using XMM-Newton observations of the source. The team also worked on data from 2010, adding new outburst times to this list. However, the new pulse times and frequencies detected for the evolution of the spin period over time are important for a better understanding of the source. We also show the long-term light curve of \sax~from The Burst Alert Telescope (BAT), an instrument of Neil Gehrels Swift Satellite with the time of \ac{RXTE} observations used in this study (brown downwards triangles) in \autoref{fig:sax-lc}.

In this study, we analyzed the temporal analysis we performed on all \ac{RXTE}\slash\ac{PCA} event data for three different intermittent-\acp{AMXP}. We present the results of our temporal analysis and describe the methods used in detail in the \autoref{temporal}. According to these results, we determined the orbital parameters for \sax~using the ML method and discussed the fact that the pulse phenomenon persists for different durations in both methods in \autoref{results}. This study contributes to the understanding of the pulse phenomenon of intermittent-\ac{AMXP} systems and a better understanding of the properties of these systems. In the end, we discussed and concluded our present results of the timing analysis in \autoref{summary}. 


\section{Temporal Analysis}
\label{temporal}

\subsection{RXTE Observations and
Data Reduction}

Temporal analysis is particularly important for determining the physical properties of sources with periodic variations.
RXTE/PCA is very effective detector with its large effective area of 6500~cm$^2$ and very high time resolution up to 1~$\mu$s.


We first obtained all available RXTE/PCA data for intermittent-AMXP; \aql, \hete~and \sax~from NASA 
archive\footnote{\href{https://heasarc.gsfc.nasa.gov/docs/archive.html}{https://heasarc.gsfc.nasa.gov/docs/archive.html}}.

The decoded data
(see \textit{decodeevt}\footnote{\hyperlink{https://heasarc.gsfc.nasa.gov/lheasoft/ftools/fhelp/decodeevt.html}{https://heasarc.gsfc.nasa.gov/lheasoft/ftools/fhelp/decodeevt.html}}) in the archive has been used.
The sample set consist of 697 SE (Science Event) and 142 GX (Good Xenon) for \aql, 
452 SE and 1 GX for \hete, while 257 SE and 11 GX for \sax.  In \autoref{table1} we present a brief information about the sources and the data. The barycentric correction to photon arrival times was performed using \textit{fxbary}\footnote{\hyperlink{https://heasarc.gsfc.nasa.gov/lheasoft/ftools/fhelp/faxbary.txt}{https://heasarc.gsfc.nasa.gov/lheasoft/ftools/fhelp/faxbary.txt}} task of HEASOFT/\textit{ftools}.

\begin{table}[!h]
    \caption{Spin and orbital information of the sources obtained from literature with the number of available RXTE data.}
    \centering
    \begin{footnotesize}
    \begin{tabular}{llll}
    \hline 
    Source  & Burst oscillation  &  Orbital Period    & \# of RXTE\\
             &  frequency $\nu$ (Hz) &   $P_{orb}$ (Hr) & Data\\
    \hline
    \aql & 550$^{1}$ & 19$^{2}$ & 839\\  
    \hete & 337$^{3}$ & 1.4$^{4}$ &  453\\ 
    \sax & 442$^{5}$ & 8.8$^{5}$ &  268\\
     
    \hline
    \end{tabular}
    \begin{flushleft}
    \vspace*{0.1cm}
    $^{1}$\citet{Casella2008}; $^{2}$\citet{1991A&A...251L..11C} $^{3}$\citet{2005ATel..516....1V}; $^{4}$\citet{2006ApJ...638..963K};  $^{5}$\citet{AltamiranoSax2008};   
    \end{flushleft}
        \end{footnotesize}
    \label{table1}
\end{table}



\subsection{$ Z_n^2$ Technique}
\label{z2technique}

The first method we used for temporal analysis is the $Z_n^2$ statistic 
\citep{1983A&A...128..245B}. The main advantage of the method is that, unlike the Fourier 
technique, it does not necessary to create a light curve and can be applied 
directly to the photon times reaching the detector. 

First, the phase values corresponding to each arrival time are calculated via; 
\begin{equation}
\Phi(t)=2\pi.\nu.t
\label{equ_phi}
\end{equation}
where $t$ is the photon arrival time and $\nu$ is the input frequency for the pulse scan. After calculating the 
phases for each photon arrival time, the $ Z_n^2$ power values for corresponding frequencies are obtained by 
using;
\begin{equation}
 Z_n^2=\frac{2}{N} \sum_{\rm k=1}^n\left[\left(\sum_{\rm j=1}^N \cos k \Phi_{\rm j}\right)^2+\left(\sum_{\rm j=1}^N \sin k \Phi_{\rm j}\right)^2\right]
\label{equ:z2}
\end{equation}
where \emph{N} is the total number of photons in the relevant time interval, \emph{n} is sinusoidal harmonics of 
increasing order and \emph{$\Phi$} is the spin phase obtained via \autoref{equ_phi}. All pulse scan has been done by using 
the single harmonic (n=1) in which the $Z_1^2$ statistic is also referred to a the Rayleigh test. 

\begin{figure}[h]
  \centering
    \includegraphics[width=3.90in, height=3.76in]{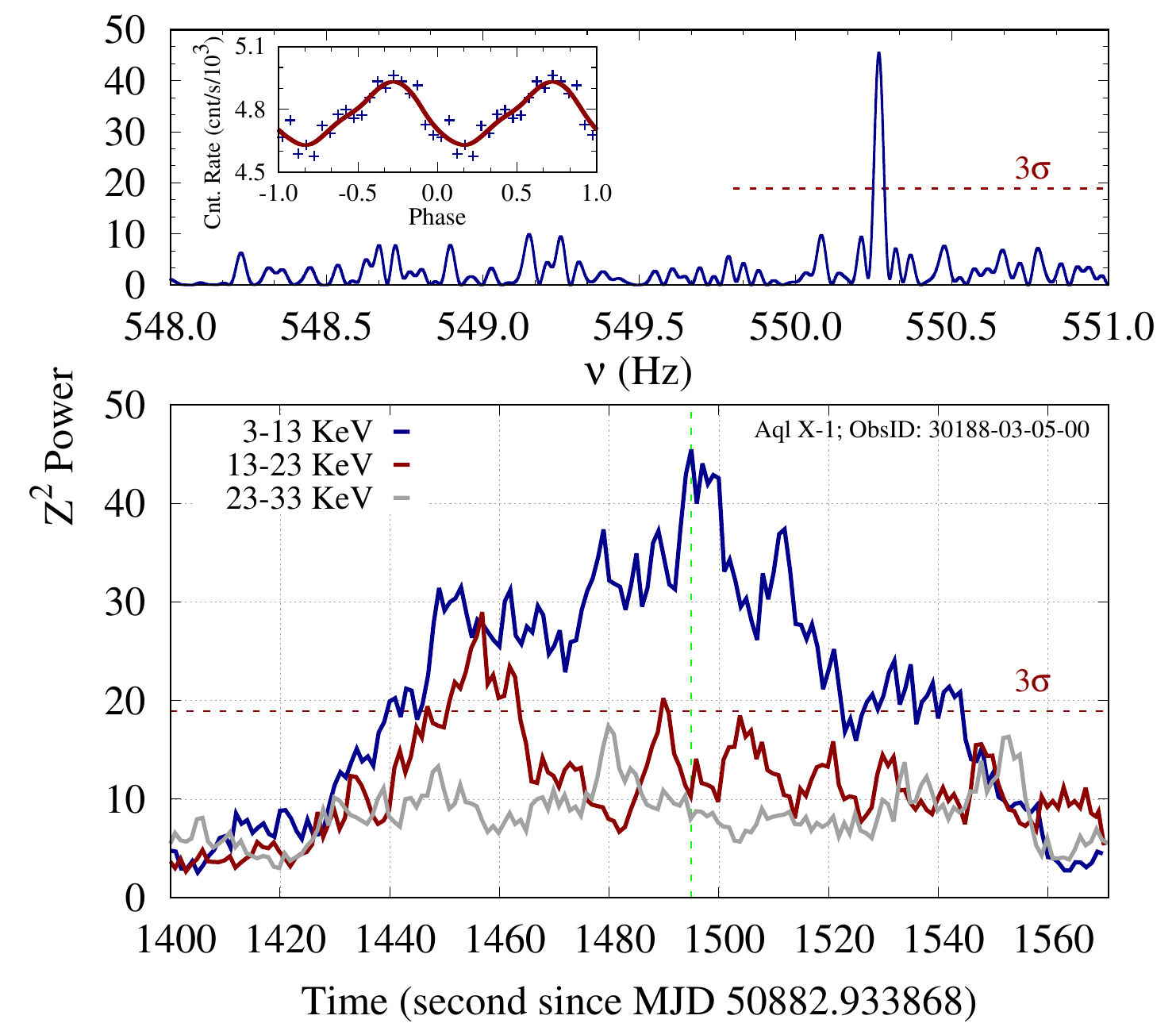}
  \caption{The power spectrum of last 150~s of the data with the ObsID of 30188-03-05-00 of 
  of \aql~(upper panel) with the pulse profile (inner panel) and the time evolution of Z$_n^2$ power (bottom) for
   energy ranges of 3.0-13.0~keV (blue), 13.0-23.0~keV (red) \& 23.0-33.0~keV (grey).}
  \label{fig:pow_energy}
\end{figure}

To examine the energy range in which the pulse is the most powerful, we first achieved the power spectrum for 
the pulse region mentioned in the literature for \aql~\ (ObsID: 30188-03-05-00; \citep{Casella2008}) around 
the pulse frequency given in \autoref{table1} (upper panel of \autoref{fig:pow_energy}) 
and repeat the scan in 3 
different energy ranges --3.0-13.0~keV, 13.0-23.0~keV, and 23.0-33.0~keV-- for the strongest pulse frequency
segmenting the data set into 25~s time intervals with 1~s shifts.

As can be seen from the \autoref{fig:pow_energy}, at the beginning of the intermittent pulse phenomena
the time evolution of the Z$_1^2$ powers of the first two energy ranges, 3.0-13.0~keV, 13.0-23.0~keV,
follow the same pattern increasing for almost $\sim$35~s. Then the first continues to increase while the 
second decrease to the same level of the third energy range, 23.0-33.0~keV. 
We followed the same procedure for \sax~whose output indicates the same.
The overall result suggests 
that it would be advantageous to perform all pulse scans of whole sample set for three intermittent 
sources, in the 3.0-13.0~keV energy range. 

After the most effective energy range was determined, by taking the pulse frequencies collected from literature as references 
for each source, we decided the pulse frequency ranges and the pulse steps to be used in the search to be able to track the 
changes in the obtained frequencies due to the orbital delay of the systems.
We performed all pulse scan in the frequency 
range of 550.0$-$550.5~Hz, 336.0$-$338.0~Hz and 441.0$-$443.0~Hz 
for \aql, \hete~and \sax, respectively with the frequency 
step of 10$^{-4}$ Hz. 

We generated power spectra via Z$_1^2$ method for all 25~s time segments continuously with 1~s shifts.
Thereby, we obtained the most powerful pulse frequency as well as its Z$_1^2$ power in the unit of 
the standard deviation for the corresponding segment.
The mean of all Z$_1^2$ power values for 25~s time intervals are 7.74, 7.17 and 6.83 while the standard deviations 
are 3.07, 4.95 and 4.95 for \aql, \hete~and \sax, respectively. Therefore, we defined the common
semi-arbitrary threshold value of Z$_1^2$ power to be 25.0
($\sim$5.6$\sigma$, $\sim$3.6$\sigma$ and $\sim$3.7$\sigma$ for each source)
to flag an interval as pulse candidate after the first elimination.

As a second elimination to exclude the false detections, we constructed the normalized pulse profiles
for all candidate in which the amplitudes vary from 0 to 1. Even if the pulse profiles are expected to follow a sinusoidal 
pattern for such sources with low magnetic field, we smoothed the pulse profiles by using s'b\'ezier method instead of 
fitting them with a sinusoidal function to take into account possible pulse profile diversity.
Root mean square (RMS) values are calculated via the discrepancies of the data in pulse profile with the smoothed curve.
Using the time segments with pulsation already known in the literature,
we defined the upper limit of RMS value as 10$^{-2}$ to flag a detection as pulse candidate.
The pulse detections with lower RMS values have smooth more likely sinusoidal patterns.
For \aql, out of 111 observations from different ObsIDs, 45 pulse detections remained after the RMS elimination.
For \sax, this number dropped from 123 to 21, while for \hete, 
30 obsID remained from a total of 244 possbile pulse candidate passing the Z$_1^2$ scan.
In addition, for each detection, single trial chance probabilities were calculated and 
the sigma values of these probability results were determined.
The resulting list given in \autoref{prelim_res_table} has newly discovered 
intermittent pulse regions as well as few already known in the literature.
To better visualize the regions with strong pulse, we created the dynamic power spectra.

\newgeometry{top=0.8in} 

\begin{figure}[H]
\centering
  \includegraphics[width=0.6\textwidth]{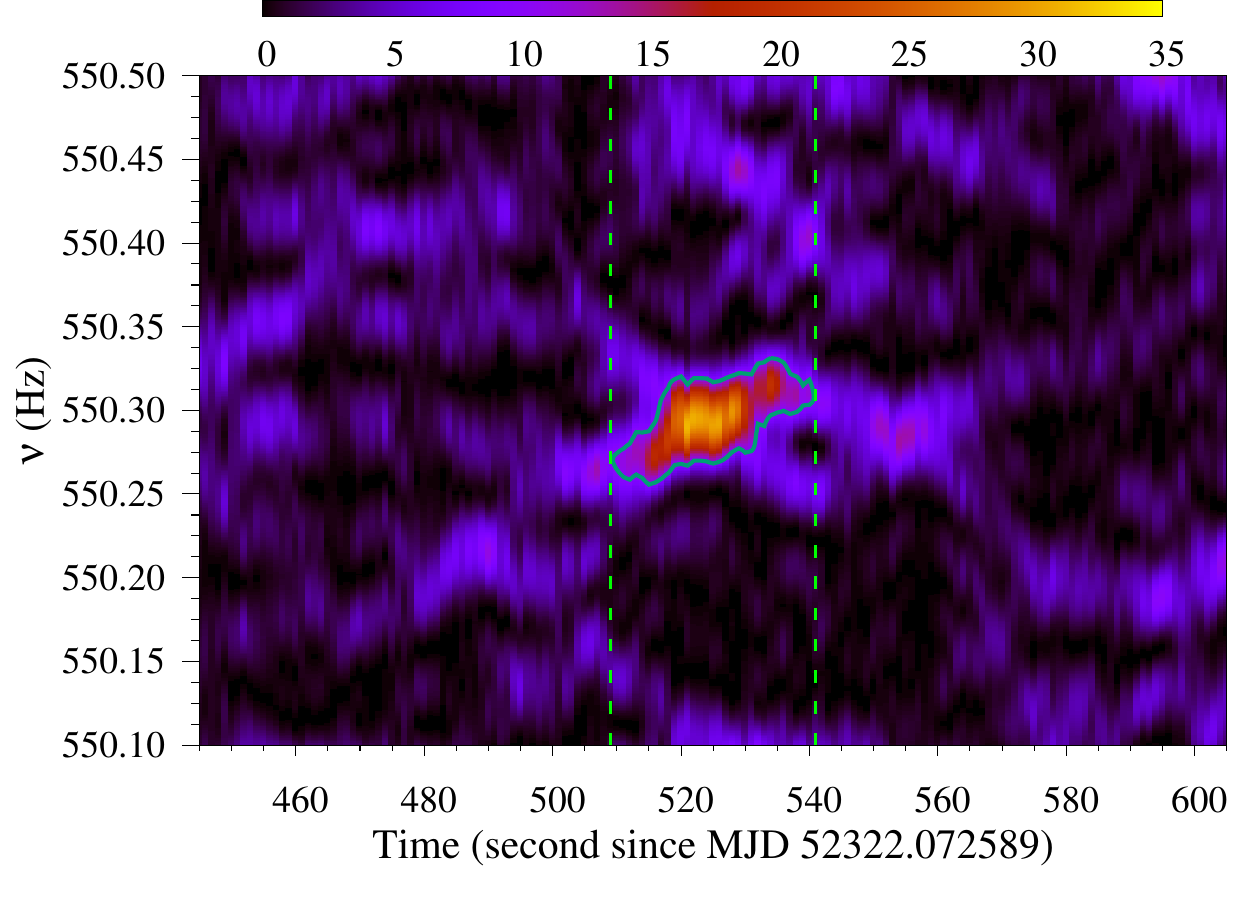}
  \includegraphics[width=0.6\textwidth]{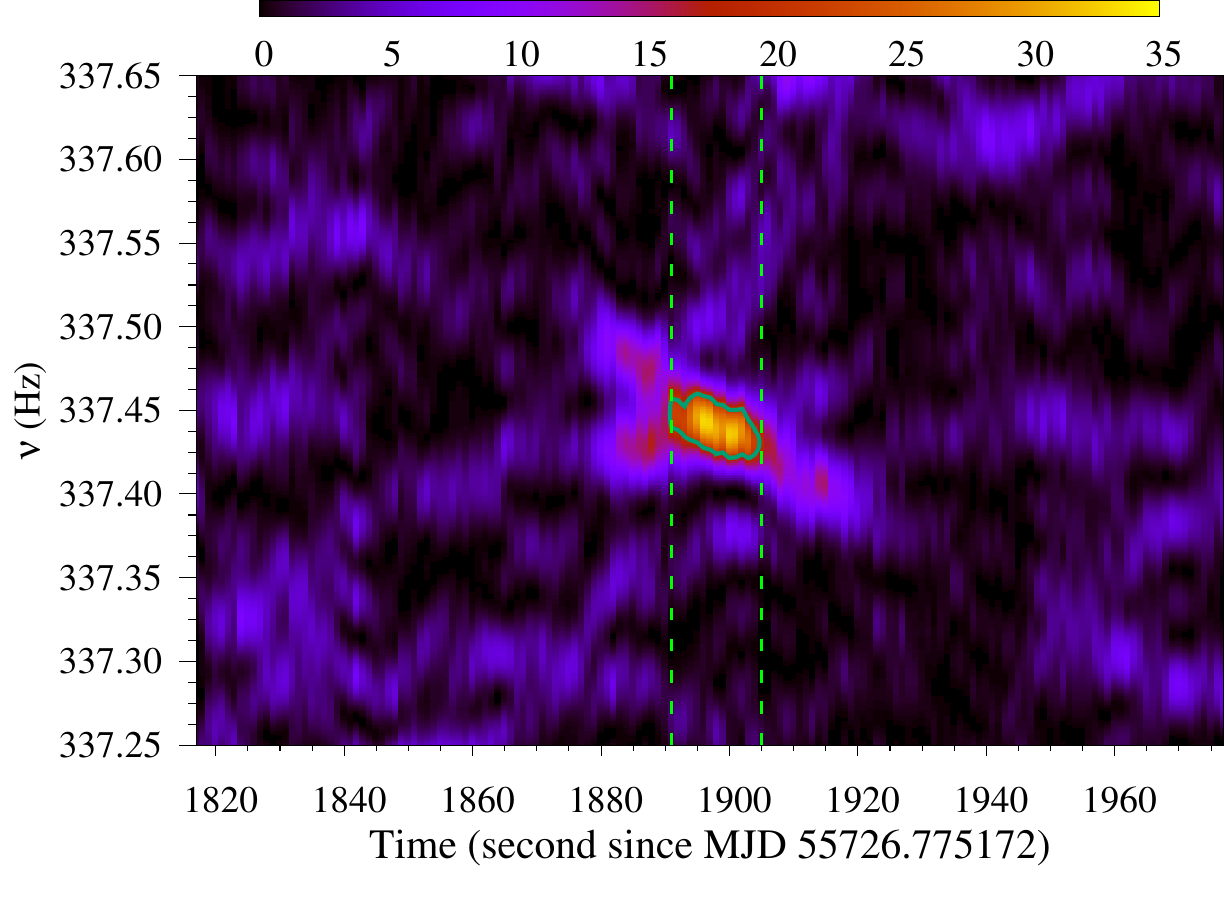}
  \includegraphics[width=0.6\textwidth]{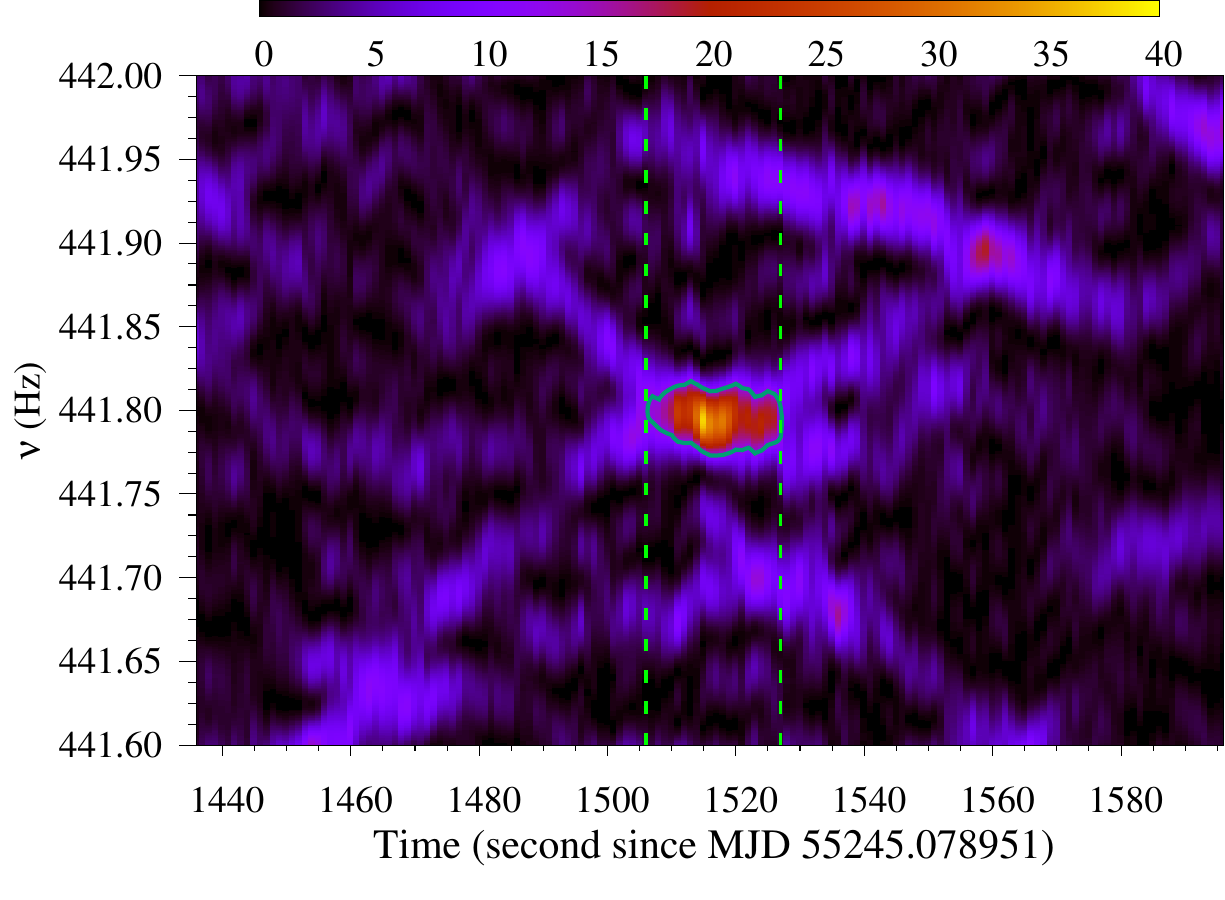}
\caption{Dynamic power spectra for selected pulse regions with high power values for \aql~(upper panel; ObsID:60429-01-03-00), 
\hete~(middle panel; ObsID: 96030-01-20-00) and \sax~(bottom panel; ObsID: 94315-01-09-11). The green contour and the green vertical lines represent the 3$\sigma$ border, the beginning and the end of the pulse-on region, respectively.}
\label{pow-all-color}
\end{figure}

\restoregeometry

\newgeometry{left=1in, right=1in, top=0.8in, bottom=1in} 

\newpage
\onecolumn
\footnotesize

\setlength{\tabcolsep}{2pt}
\begin{longtable}{ccccccclc}
\caption{Information of the detected pulse candidates.}
\label{prelim_res_table}\\
\hline
\hline
\textbf{\textit{Source}}  & \textbf{\textit{ObsID}} & \textbf{\textit{Start Time}} &  \textbf{\textit{Cnt Rate}} &  \textbf{\textit{$f_s$}} & \textbf{\textit{$Z^{2}$}}   & \textbf{\textit{Pulse}} & \textbf{\textit{Prob}} & \textbf{\textit{$\sigma^{\ast}_{Prob}$}}\\

         &  & \textbf{\textit{(MJD)}}  & \textbf{\textit{(cnt/sec)}} & \textbf{\textit{(Hz)}} &\textbf{\textit{Power}}  & \textbf{\textit{dur.(s)}} & \textit{\textbf{(Trial)}}  &  
    \\\hline

\\[-0.9em]
{\multirow{6}{*}{\rotatebox[origin=c]{90}{\normalsize \textbf{\aql}}}}

&20098-03-01-00 & 50496.10567877 & 2688 & 550.1650 & 28.19 & 28 &  $3.78 \times 10^{-3}$ & 2.90 \\
&20092-01-04-00 & 50692.28575460 & 190 & 550.2660 & 28.83 & 28 &  $2.75 \times 10^{-3}$ & 2.99 \\
&20092-01-04-02 & 50694.08610572 &  562 & 550.1035 & 27.45 & 26 &  $5.47 \times 10^{-3}$ & 2.78 \\
&20092-01-05-01 & 50697.60477778 & 1461 & 550.4900 & 27.61 & 28 &  $5.05 \times 10^{-3}$ & 2.80 \\
&20092-01-05-08 & 50701.73715716 & 1747 & 550.2254 & 28.19 & 25 &  $3.78 \times 10^{-3}$ & 2.90 \\
&20092-01-05-09 & 50701.99710234 & 1717 & 550.0658 & 28.22 & 27 &  $3.72 \times 10^{-3}$ & 2.90 \\
&30072-01-01-02 & 50876.58891642 & 2020 & 550.2615 & 27.65 & 26 &  $4.95 \times 10^{-3}$ & 2.81 \\
&30188-03-05-00 & 50882.94993491 & 4956 & 550.2751 & 42.86 & 109 &  $2.47 \times 10^{-6}$ & 4.71 \\
&40047-01-01-02 & 51319.53605762 & 3720 & 550.2080 & 28.22 & 26 &  $3.72 \times 10^{-3}$ & 2.90 \\
&40047-02-03-00 & 51322.82465716 & 2857 & 550.4879 & 27.22 & 25 &  $6.14 \times 10^{-3}$ & 2.74 \\
&40047-03-02-00 & 51332.81198419 & 1044 & 550.0962 & 27.37 & 28 &  $5.70 \times 10^{-3}$ & 2.76 \\
&40047-03-08-00 & 51338.73194882 & 297 & 550.3678 & 27.36 & 25 &  $5.73 \times 10^{-3}$ & 2.76 \\
&40033-10-02-00 & 51396.73089558 & 387 & 550.2904 & 27.08 & 25 &  $6.59 \times 10^{-3}$ & 2.72 \\
&40048-01-04-00 & 51416.51510855 & 224 & 550.4761 & 28.09 & 28 &  $3.97 \times 10^{-3}$ & 2.88 \\
&40048-01-09-01 & 51439.77484334 & 140 & 550.1436 & 28.77 & 26 &  $2.83 \times 10^{-3}$ & 2.99 \\
&40432-01-04-00 & 51471.31175207 & 332 & 550.4187 & 30.53 & 33 &  $1.17 \times 10^{-3}$ & 3.25 \\
&40432-01-05-00 & 51478.30538632 & 242 & 550.1763 & 29.73 & 26 &  $1.75 \times 10^{-3}$ & 3.13 \\
&50049-02-04-00 & 51834.41329142 & 5126 & 550.0919 & 30.87 & 39 &  $9.90 \times 10^{-4}$ & 3.29 \\
&50049-02-07-04 & 51843.44083771 & 4376 & 550.3557 & 27.21 & 25 &  $6.17 \times 10^{-3}$ & 2.74 \\
&50049-02-11-00 & 51851.24597660 & 5206 & 550.2675 & 27.37 & 25 &  $5.70 \times 10^{-3}$ & 2.76 \\
&50049-02-13-00 & 51855.31337750 & 1786 & 550.0654 & 33.38 & 36 &  $2.82 \times 10^{-4}$ & 3.63 \\
&50049-02-13-01 & 51856.15405906 & 3448 & 550.2060 & 25.69 & 26 &  $1.32 \times 10^{-2}$ & 2.48 \\
&50049-02-15-00 & 51859.29498123 & 2514 & 550.2466 & 29.52 & 26 &  $1.94 \times 10^{-3}$ & 3.10 \\
&50049-02-15-07 & 51864.27405531 & 640 & 550.4536 & 27.92 & 25 &  $4.33 \times 10^{-3}$ & 2.85 \\
&50049-03-04-00 & 51869.50632613 & 165 & 550.3728 & 29.09 & 28 &  $2.41 \times 10^{-3}$ & 3.03 \\
&50049-03-05-00 & 51870.36583771 & 152 & 550.2448 & 27.35 & 25 &  $5.75 \times 10^{-3}$ & 2.76 \\
&60054-02-01-05 & 52088.68913504 & 349 & 550.2891 & 27.31 & 29 &  $5.87 \times 10^{-3}$ & 2.75 \\
&60054-02-04-01 & 52104.99310723 & 208 & 550.3810 & 27.84 & 28 &  $4.50 \times 10^{-3}$ & 2.84 \\
&60054-02-04-05 & 52109.28628910 & 156 & 550.2271 & 27.84 & 26 &  $4.50 \times 10^{-3}$ & 2.84 \\
&60429-01-03-00 & 52322.07790389 & 1109 & 550.2938 & 30.16 & 35 &  $1.41 \times 10^{-3}$ & 3.19 \\
&70069-03-03-13 & 52354.11222252 & 781 & 550.4317 & 27.95 & 27 &  $4.26 \times 10^{-3}$ & 2.86 \\
&70426-01-01-00 & 52695.85008540 & 594 & 550.4656 & 28.35 & 25 &  $3.49 \times 10^{-3}$ & 2.92 \\
&90403-01-02-00 & 53069.98169419 & 534 & 550.1071 & 27.08 & 25 &  $6.59 \times 10^{-3}$ & 2.72 \\
&90017-01-04-00 & 53150.94205780 & 694 & 550.1979 & 26.82 & 26 &  $6.96 \times 10^{-3}$ & 2.68 \\
&90017-01-06-00 & 53153.81474512 &  847 & 550.2747 & 28.05 & 28 &  $3.93 \times 10^{-3}$ & 2.87 \\

\\[-0.9em]
\hline
\multirow{12}{*}{\rotatebox[origin=c]{90}{\normalsize \textbf{\hete}}}
\\[-0.9em]

&91015-01-03-00 & 53538.76905229 & 257 & 337.8941 & 33.13 & 44 & $1.28 \times 10^{-3}$ & 3.22 \\
&91015-01-03-01 & 53539.81025151 & 273 & 336.9985 & 30.84 & 35 & $4.02 \times 10^{-3}$ & 2.88 \\
&91015-01-03-03 & 53541.79347351 & 117 & 336.1608 & 33.46 & 33 & $1.08 \times 10^{-3}$ & 3.27 \\
&91015-01-05-00 & 53553.64309161 & 273 & 337.0225 & 30.31 & 30 & $5.24 \times 10^{-3}$ & 2.79 \\
&91059-03-01-04 & 53572.98533258 & 291 & 336.7594 & 31.28 & 38 & $3.23 \times 10^{-3}$ & 2.95 \\
&91059-03-02-00 & 53573.83126993 & 295 & 337.1900 & 31.48 & 35 & $2.92 \times 10^{-3}$ & 2.98 \\
&91059-03-02-01 & 53574.63439556 & 186 & 337.4440 & 32.23 & 29 & $2.01 \times 10^{-3}$ & 3.09 \\
&91057-01-01-01 & 53592.57316085 & 206 & 336.7003 & 33.06 & 33 & $1.32 \times 10^{-3}$ & 3.21 \\
&91057-01-04-02 & 53611.07035818 & 421 & 337.3970 & 30.44 & 35 & $4.91 \times 10^{-3}$ & 2.81 \\
&91057-01-05-01 & 53617.59555207 & 243 & 336.7877 & 30.20 & 39 & $5.54 \times 10^{-3}$ & 2.77 \\
&92049-01-23-00 & 53927.87756866 & 304 & 337.0165 & 31.19 & 27 & $3.37 \times 10^{-3}$ & 2.93 \\
&92049-01-35-00 & 54009.40736580 & 299 & 336.8748 & 30.83 & 38 & $4.04 \times 10^{-3}$ & 2.88 \\
&92049-01-47-00 & 54150.06003732 & 238 & 336.7169 & 30.27 & 29 & $5.35 \times 10^{-3}$ & 2.79 \\
&92049-01-59-00 & 54234.69700436 & 201 & 336.6769 & 30.03 & 44 & $6.03 \times 10^{-3}$ & 2.75 \\
&93030-01-21-00 & 54424.36103819 & 103 & 336.0345 & 30.14 & 27 & $5.70 \times 10^{-3}$ & 2.76 \\
&93030-01-23-00 & 54439.22827984 & 160 & 337.0374 & 30.77 & 31 & $4.16 \times 10^{-3}$ & 2.87 \\
&93030-01-40-00 & 54610.69024152 & 71 & 336.9616 & 32.83 & 31 & $1.49 \times 10^{-3}$ & 3.18 \\
&93030-01-50-00 & 54680.85659227 & 329 & 337.1594 & 31.01 & 40 & $3.69 \times 10^{-3}$ & 2.90 \\
&93451-01-01-00 & 54752.43204130 & 191 & 337.4422 & 31.32 & 27 & $3.16 \times 10^{-3}$ & 2.95 \\
&93451-01-02-00 & 54760.02068789 & 174 & 337.2471 & 31.87 & 37 & $2.40 \times 10^{-3}$ & 3.04 \\
&94030-01-10-00 & 54925.47132619 & 183 & 336.2797 & 30.62 & 29 & $4.49 \times 10^{-3}$ & 2.84 \\
&94030-01-45-00 & 55170.02184756 & 124 & 336.3612 & 31.92 & 62 & $2.34 \times 10^{-3}$ & 3.04 \\
&95030-01-15-00 & 55328.61154332 & 200 & 336.7656 & 33.56 & 33 & $1.03 \times 10^{-3}$ & 3.28 \\
&95030-01-18-00 & 55346.71370884 & 33 & 337.0820 & 32.22 & 32 & $2.02 \times 10^{-3}$ & 3.09 \\
&95030-01-21-00 & 55370.44460938 & 55 & 337.3743 & 31.86 & 28 & $2.41 \times 10^{-3}$ & 3.03 \\
&95030-01-22-00 & 55374.48661165 & 67 & 337.7756 & 33.53 & 27 & $1.05 \times 10^{-3}$ & 3.28 \\
&95030-01-39-00 & 55494.37598280 & 155 & 336.0014 & 30.38 & 33 & $5.06 \times 10^{-3}$ & 2.80 \\
&96030-01-14-00 & 55685.37969921 & 213 & 337.0181 & 31.41 & 32 & $3.02 \times 10^{-3}$ & 2.97 \\
&96030-01-20-00 & 55726.79637844 & 232 & 337.4403 & 31.82 & 34 & $2.46 \times 10^{-3}$ & 3.03 \\
&96030-01-24-00 & 55755.72228514 & 201 & 336.8407 & 32.22 & 32 & $2.02 \times 10^{-3}$ & 3.09 \\

 \\[-0.9em]
\hline
{\multirow{11}{*}{\rotatebox[origin=c]{90}{\normalsize \textbf{\sax}}}}
\\[-0.9em]

 & 60035-02-02-04 & 52191.66631225              & 1311  & 442.3835 & 31.24 & 27 &  $3.28 \times 10^{-3}$ & 2.94 \\
 & 60035-02-02-05$^\dagger$ & 52192.31714558    & 1506  & 442.3512 & 36.14 & 36 & $2.84 \times 10^{-4}$ & 3.63 \\
 & 60035-02-03-01 & 52195.29753910              & 1342   & 442.8277 & 38.68 & 38 & $7.99 \times 10^{-5}$ & 3.94 \\
 & 60035-02-03-00$^\ddagger$ & 52195.52094188   & 1344 & 442.3241 & 34.14 & 97 & $7.70 \times 10^{-4}$ & 3.36 \\
 & 60035-02-03-00 & 52195.56367336              & 1445 & 442.3424 & 29.13 & 46 & $9.46 \times 10^{-3}$ & 2.59 \\
 & 60035-02-03-00 & 52195.56912475              & 1465 & 441.9936 & 31.00 & 28 &  $3.72 \times 10^{-3}$ & 2.90 \\
 & 60035-02-03-02 & 52198.21734234              & 1129 & 442.9103 & 31.11 & 27 &  $3.51 \times 10^{-3}$ & 2.92 \\
 & 91050-03-07-00 & 53535.46242052              & 1008 & 442.3380 & 32.78 & 661 & $1.53 \times 10^{-3}$ & 3.17 \\
 & 94315-01-05-01 & 55214.82650901              & 1542 & 442.4821 & 29.50 & 31 & $7.86 \times 10^{-3}$ & 2.66 \\
 & 94315-01-05-02$^\dagger$ & 55215.98128910    & 275 & 442.2458 & 30.00 & 31 & $6.11 \times 10^{-3}$ & 2.74 \\
 & 94315-01-05-03$^\dagger$ & 55216.77845345    & 683 & 441.7748 & 30.51 & 30 & $4.74 \times 10^{-3}$ & 2.82 \\
 & 94315-01-06-07 & 55222.54533346              & 824 & 442.3521 & 35.16 & 99 & $4.64 \times 10^{-4}$ & 3.50 \\
 & 94315-01-07-01 & 55225.68381963              & 629 & 441.0436 & 29.56 & 35 & $7.62 \times 10^{-3}$ & 2.67 \\
 & 94315-01-07-05 & 55225.78354605              & 608 & 442.6948 & 31.94 & 26 & $2.32 \times 10^{-3}$ & 3.05 \\
 & 94315-01-07-03 & 55231.44675207              & 210 & 442.2866 & 30.10 & 26 & $5.82 \times 10^{-3}$ & 2.76 \\
 & 94315-01-08-00 & 55232.85877785              & 358 & 442.1112 & 30.48 & 31 & $4.80 \times 10^{-3}$ & 2.82 \\
 & 94315-01-08-04 & 55235.10170761              & 250 & 441.2279 & 29.73 & 37 & $7.00 \times 10^{-3}$ & 2.70 \\
 & 94315-01-08-05 & 55236.48482535              & 174 & 441.2002 & 29.11 & 31 & $9.55 \times 10^{-3}$ & 2.59 \\
 & 94315-01-08-06 & 55237.45353447              & 147 & 442.9231 & 29.40 & 32 & $8.30 \times 10^{-3}$ & 2.64 \\
 & 94315-01-09-11 & 55245.09670577              & 167 & 441.7932 & 37.52 & 30 & $1.43 \times 10^{-4}$ & 3.80 \\
 & 94315-01-27-00 & 55487.77792105              & 32 & 441.0142 & 32.54 & 27 & $1.72 \times 10^{-3}$ & 3.13 \\
\\[-0.9em]
\hline
\hline

\end{longtable}

 \begin{flushleft}
    \vspace*{0.1cm}
$^\ast$These values are the sigmas obtained for the single trial probability    

$^\dagger$\aql; ObsID: 90017-01-06-00; the pulse is detected 42.5~s before a thermonuclear X-ray burst.
\sax;ObsID: 60035-02-02-05, the pulse is detected 222.1~s before and ObsIDs: 60035-02-02-05 and 94315-01-05-03, 
the pulses are detected 387.6~s and 207.5~s after the thermonuclear X-ray burst, respectively.

$^\ddagger$The pulse detection is used to get the model to be used in ML method which provides the orbital period of \sax.
    \end{flushleft}

\normalsize

\restoregeometry

We show an example dynamic power spectrum with high power values and relatively short durations 
for each intermittent-AMXP in \autoref{pow-all-color}.
The green contour indicates the 3$\sigma$ border around the pulse region. 
In addition, green vertical lines are added to show the beginning and end of the pulse active area. 

Finally, the new detected pulse frequencies and the pulse profiles created based on this
frequency are prepared to be used as input parameters for the ML technique.


\subsection{Maximum Likelihood Technique}

The \ac{ML}  technique, first proposed by \citet{2009ApJ...706.1163L} and later 
improved by \citet{2014ApJ...783...99S}, is used to 
determine the spin phase shift based on a given pulse profile model.
Therefore, the handicap of the method is the necessary of a model which will come from previous pulse detection. In our search, 
we use the  pulse detections from Z$_1^2$ method as an input. The methodology of the ML technique is described below via the 
detected pulse region of \aql~which was given in \autoref{z2technique}.

\textit{(i)} First, Spin phases, $\phi(T_i)$, for all barycentering corrected photon arrival times (T$_i$) 
within the corresponding time segment are calculated via
\begin{equation}
\phi_i(T_i) = \nu(T_i - T_0) - \lfloor\nu(T_i - T_0)\rfloor
\end{equation}
where T$_0$ is the epoch and $\nu$ is the spin frequency which are obtained 
from the model of detected pulse regions via Z$_1^2$ method.

\textit{(ii)} The $\phi_i$ values ($0 \leq \phi_i < 1$) are then grouped with the selected phase width.
The resulting histogram depicts the pulse profile of the source.
The pulse profile must be described either as a continuous or as ordered data with small intervals.
Therefore, it's essential to model the pulse profile using a function or any smoothing technique. 
In \autoref{fig:pow_energy}, we show the pulse profile and the smoothed model using S-Bézier method from the 
last 150 s time interval of the RXTE data (ObsID: 30188-03-05-00) of \aql.
We then get an array with intensity values, I$_i$, for chosen phase step, $\Delta \phi$ ($=$10$^{-4}$ for this study).

\textit{(iii)} To use the pulse profile as a probability function,
it is needed to normalize the smoothed pulse profile as the summation of all probabilities giving 100\%.
In other words, the integral of the pulse profile must be equal to 1.
Therefore, for the resulting discrete data with n sample ($n=1/\Delta \phi$), the $I_i$ values 
are converted to probability values, $P_i(\phi)$, using 
\begin{equation}
P_i(\phi) = \frac{I_i}{\Delta \phi \sum_{i=1}^{n} I_i}.
\label{eq_prob}
\end{equation}

\newgeometry{top=0.8in} 

\textit{(iv)} At this stage, we've gathered all the necessary materials to apply the ML technique. 
In the first step (\textit{i}), we have calculated the phase values associated with the photon arrival times in the dataset.
Now, the probability of a given phase shift, $\phi_\text{off}$, is calculated by using all photons in the 
corresponding time interval according to probability function obtained in the previous step via
\begin{equation}
{\text{Prob}}(\phi_{\text{off}}) = \prod_{i=1}^{N} P_i(\phi_i - \phi_{\text{off}})
\label{eq_prob2}
\end{equation}
where, N is the total number of photons in the interval, and $\phi{_i}$ is the phase values.
To obtain the P$_i(\phi)$ value for any $\phi$, it might be necessary to use an interpolation technique,
since $\phi_i$ values are obtained in a tabular form from smoothed and normalized pulse profile. 

\begin{figure}[H]
  \centering
  \includegraphics[width=3.10in, height=3.0in]{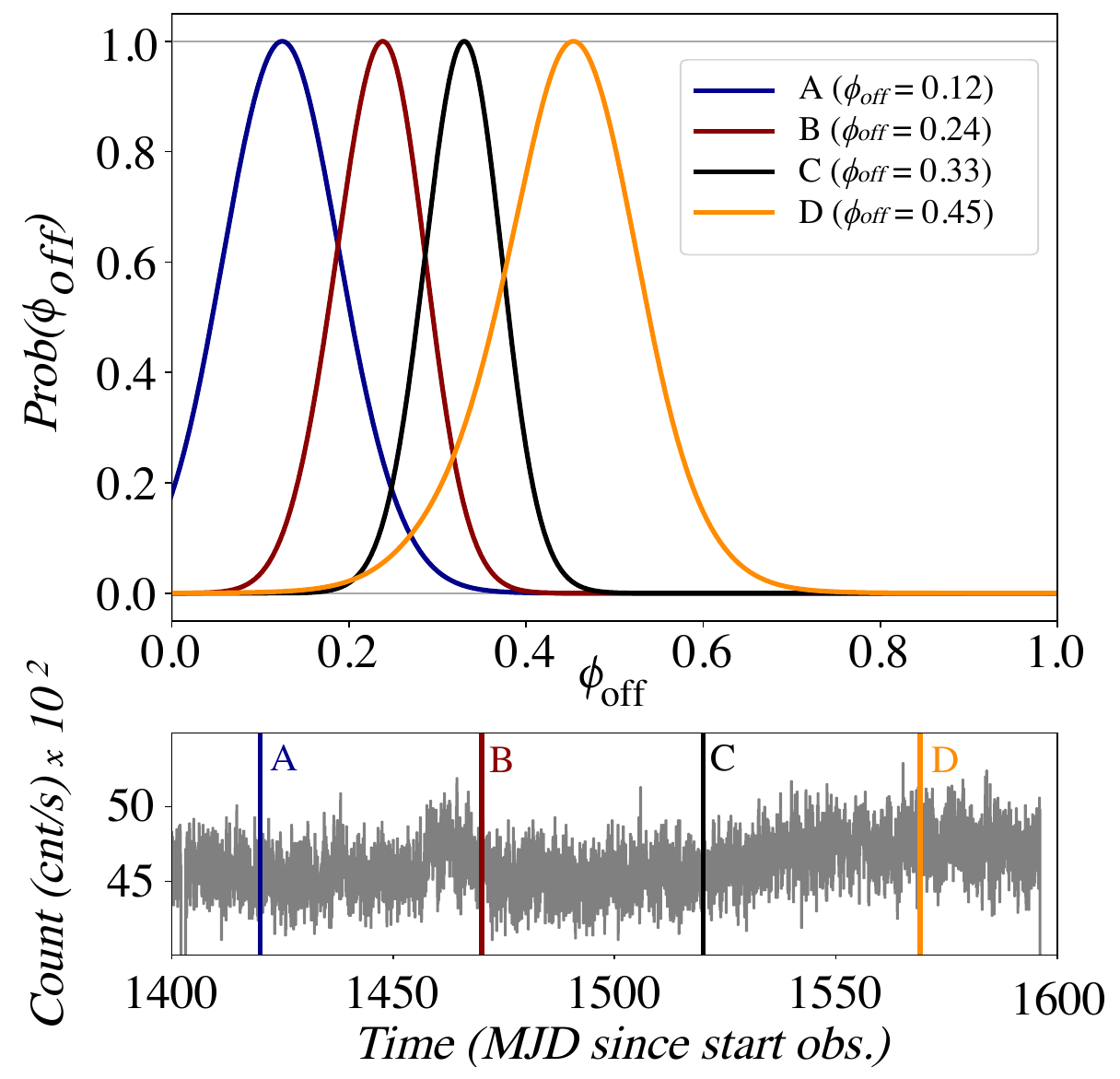}
  \caption{Phase probability distributions of the pulse profile obtained for the strongest pulse region for the \aql, shown in different time bins of 25~s each. The curves for \ac{RXTE}/\ac{PCA} data with observation number 30188-03-05-00, starting from the beginning of the observation, correspond to the following color and time intervals: blue (1420$^{th}$~s), red 
  (1470$^{th}$~s), black (1520$^{th}$~s), and orange (1570$^{th}$~s).}
  \label{fig:Prob}
\end{figure}

\textit{(v)} We, then, calculate the probability values for a set of $\phi_\text{off}$ and obtain the probability distribution for the corresponding time segment.
Subsequently, modelling the probability distribution with a Gaussian curve provides us the most probable $\phi_\text{off}$ as well as the error from the  full 
width at half maximum (FWHM).
In \autoref{fig:Prob}, we show four distinct probability distribution obtained from the successive time segments with 25~s duration of \aql~with the 
ObsID of 30188-03-05-00. It is easy to see the shift in the most probable $\phi_\text{off}$ which is mainly due to the orbital motion of the source.

\begin{figure}[H]
\centering
  \includegraphics[width=3.90in, height=2.76in]{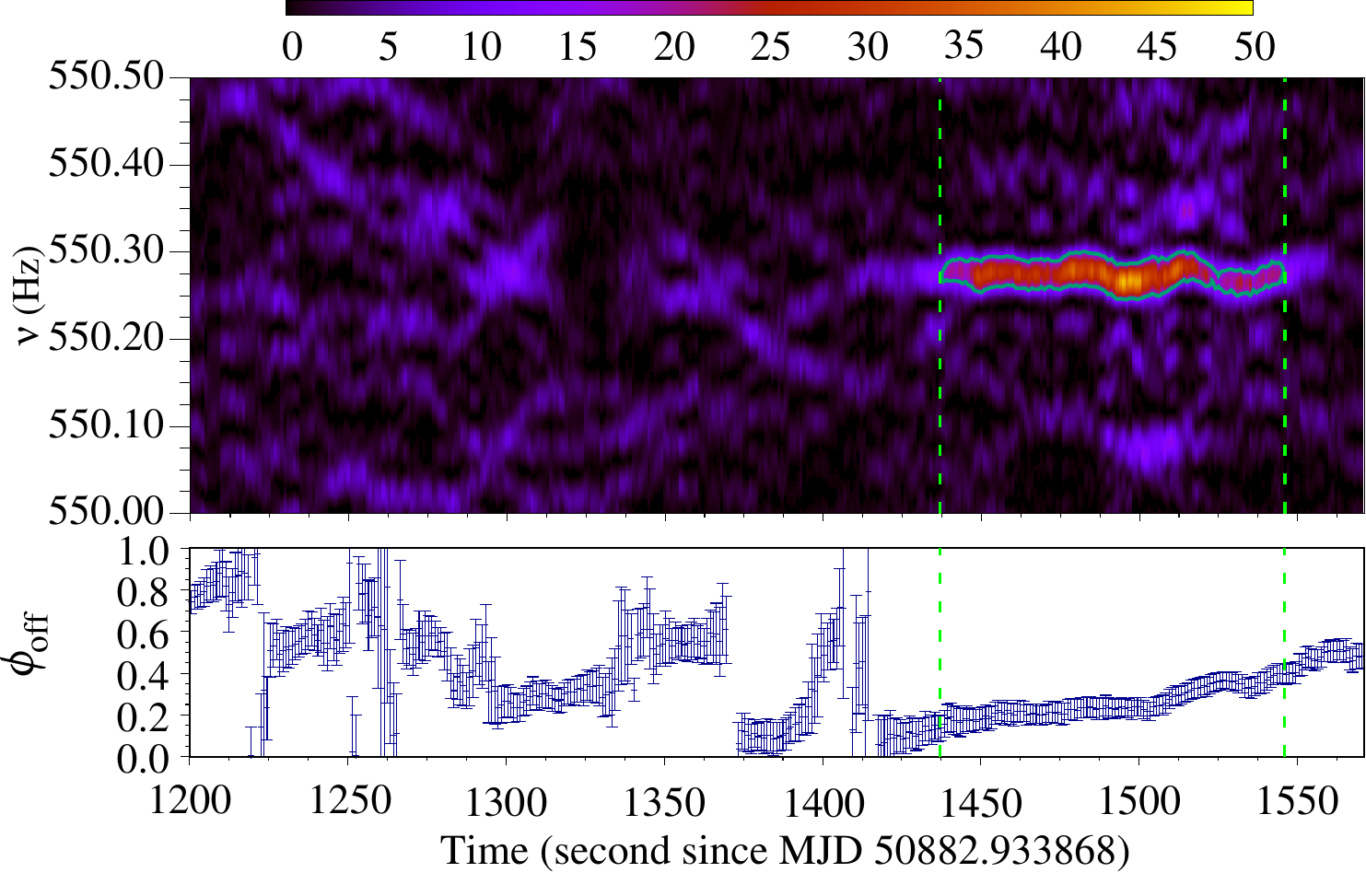}
    \includegraphics[width=3.90in, height=2.76in]{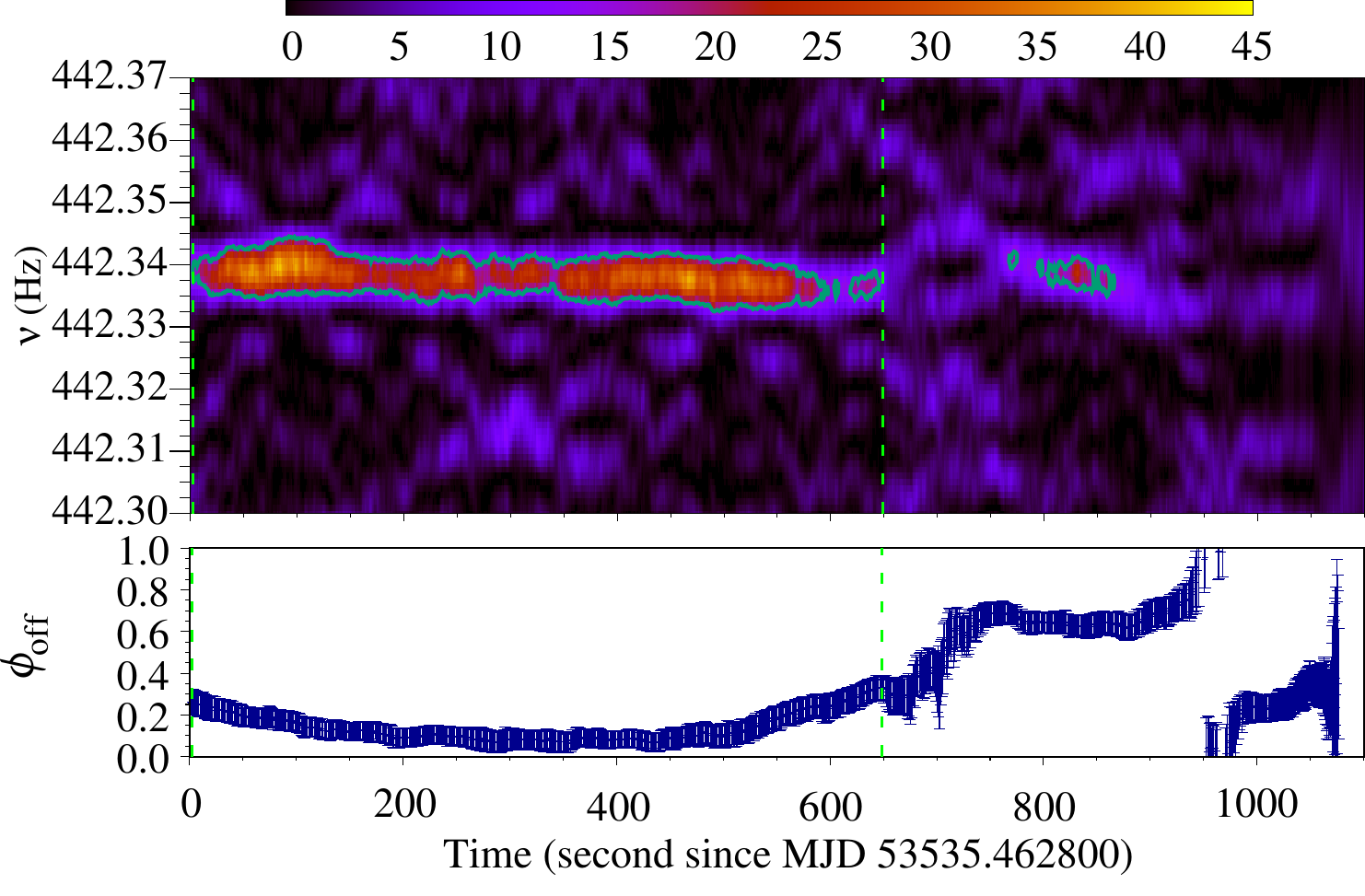}
  \caption{The dynamic power spectra of the strongest pulse regions for \aql~(ObsID: 30188-03-05-00; upper panels) and 
  \sax~(ObsID: 91050-03-07-00; bottom panels) 
  via $Z^{2}$ (top). Regions above 3~$\sigma$ are highlighted by the green contour.
  The vertical dashed green lines are to emphasize the pulse durations.
  The time evolution of the phase shifts obtained via \ac{ML} are given in the bottom panels.}
\label{fig:aql-sax-literature}
\end{figure}

In \autoref{fig:aql-sax-literature}, we show the dynamic power spectrum created with the Z$_1^2$ power values for 25~s time intervals with 1~s shifts 
for \aql~(ObsID: 30188-03-05-00) and \sax~(ObsID: 91050-03-07-00; top panels), and the most porbable $\phi_{\text{off}}$ values obtained via the ML technique
for the same time intervals (bottom panels). In the dynamic spectrum, strong pulse regions (above $3\sigma$) are depicted in contour lines, with the beginning and the terminal 
times highlighted by green vertical lines. The result of the ML which we applied to the times when the pulse was previously reported in the literature, shows 
systematic changes in the $\phi_{\text{off}}$ values.

\restoregeometry

\section{Results and Discussion}
\label{results}

For the first intermittent-\ac{AMXP} in our study, \textit{\aql}, 111 pulse candidates were detected 
based on Z$_1^2$ test in 839 RXTE/PCA data with a total exposure time of 2186 ks.
After applying RMS elimination, this number decreased to 45. 
Successive 25~s-long segment within these detections were combined,
though we announce 35 pulse candidates while the shortest one is 25~s long and the longest is 109~s.

The times of the pulse candidates are shown as black vertical lines on \ac{ASM} light curves (\autoref{fig:aql-lc}) and also the details of them are listed in \autoref{prelim_res_table}. In addition, the trial probabilities and sigma values of these probabilities are given in \autoref{prelim_res_table}.

The top panel of \autoref{fig:aql-sax-literature} presents the dynamic power spectra of Aql X-1 for the ObsID of 30188-03-05-00 where the pulse is reported in \citet{Casella2008}.  The green contour indicates the edges with the 3~$\sigma$ value and vertical green line shows the time duration based on Z$_1^2$ method. The bottom panel shows the phase offset values for consecutive 25~s time segments with 1~s shifts via ML method by using the pulse profile obtained from the last 120~s segment. The systematic change in the phase offsets is the hint for the duration of the pulse-on region via ML. Although there is no clear difference, the duration obtained via ML is slightly longer than the one obtain from the Z$_1^2$  which allows us to argue the smooth transition from pulse-on to pulse-off behavior. 

For \textit{\hete}, we report 30 pulse candidates after RMS elimination over 244 25~s-long time segments after Z$^2$
out of 453 RXTE/PCA data with an observation period of 1128~ks. The duration of the shortest detection is 27~s while the 
longest one is 62~s. The detected pule regions is shown as vertical black lines in the long term light curve 
(\autoref{fig:HETE-lc}) and the details are given in \autoref{prelim_res_table}. However, the phase offset values obtained 
via ML method are rather uncertain because the pulse durations are quite short compared to the other two sources.

For \textit{\sax}, 21 pulse candidates were identified after RMS elimination from 123 data exceeding
the Z$_1^2$ power threshold within 1426~ks in 258 RXTE/PCA data.
The shortest and longest pulse candaites are 26~s and 661~s, respectively, while the total pulse duration is 557~ks.
The details of the output of the search are again given in the last part 
of \autoref{prelim_res_table}. We give an interesting example of detected pulse candidate 
in the right panel of \autoref{fig:aql-sax-literature}. It is clearly seen 
from the figure that the pulse is invisible after $\sim$625$^{th}$~s and becomes visible again $\sim$800$^{th}$~s of the data for very short time.
The source is located in the globular cluster NGC 6440.
The long term Swift/BAT light curve of the cluster is given together with the RXTE 
data (upside-down brown triangles) and the detected pulse candidates (vertical black lines) in \autoref{fig:sax-lc}. We can argue from the long 
term light curve that the times of the detected pulse candidates have no obvious relation with the luminosity level since the source does not show 
clear luminosity change such as energetic outburst, etc.

In addition to chance probability and sigma calculations, we performed a set of simulations to check the
reliability of the pulse detections in our study.
We generated 25~s--long time series with random events for a given 
average count rate. We, then, applied the same procedure to search for pulsations via Z$_1^2$ to the simulated time segment.
We repeated this routine 10$^5$ times for the brightest and the faintest count-rated pulse candidate following the same 
criteria for each source. The artificial data with higher
Z$^2$ power values then threshold level is again marked as candidates, then RMS
elimination is applied to those data. The number of the artificial 25~s--long time series passing these
two steps provides the chance probabilities for the brightest and the faintest
pulse candidates for each source given in \autoref{prelim_res_table}. 
Throughout the 10$^5$ trial simulation, we can conclude that chance probabilities of the faintest and the brightest
pulse candidates of \aql~are 2$\times$10$^{-5}$ and 3$\times$10$^{-5}$, respectively.
These values are 1$\times$10$^{-5}$ and 3$\times$10$^{-5}$ for \hete~while 2$\times$10$^{-5}$
and 4$\times$10$^{-5}$ for \sax.

\begin{figure}[H]
\centering
  \includegraphics[width=3.90in, height=2.76in]{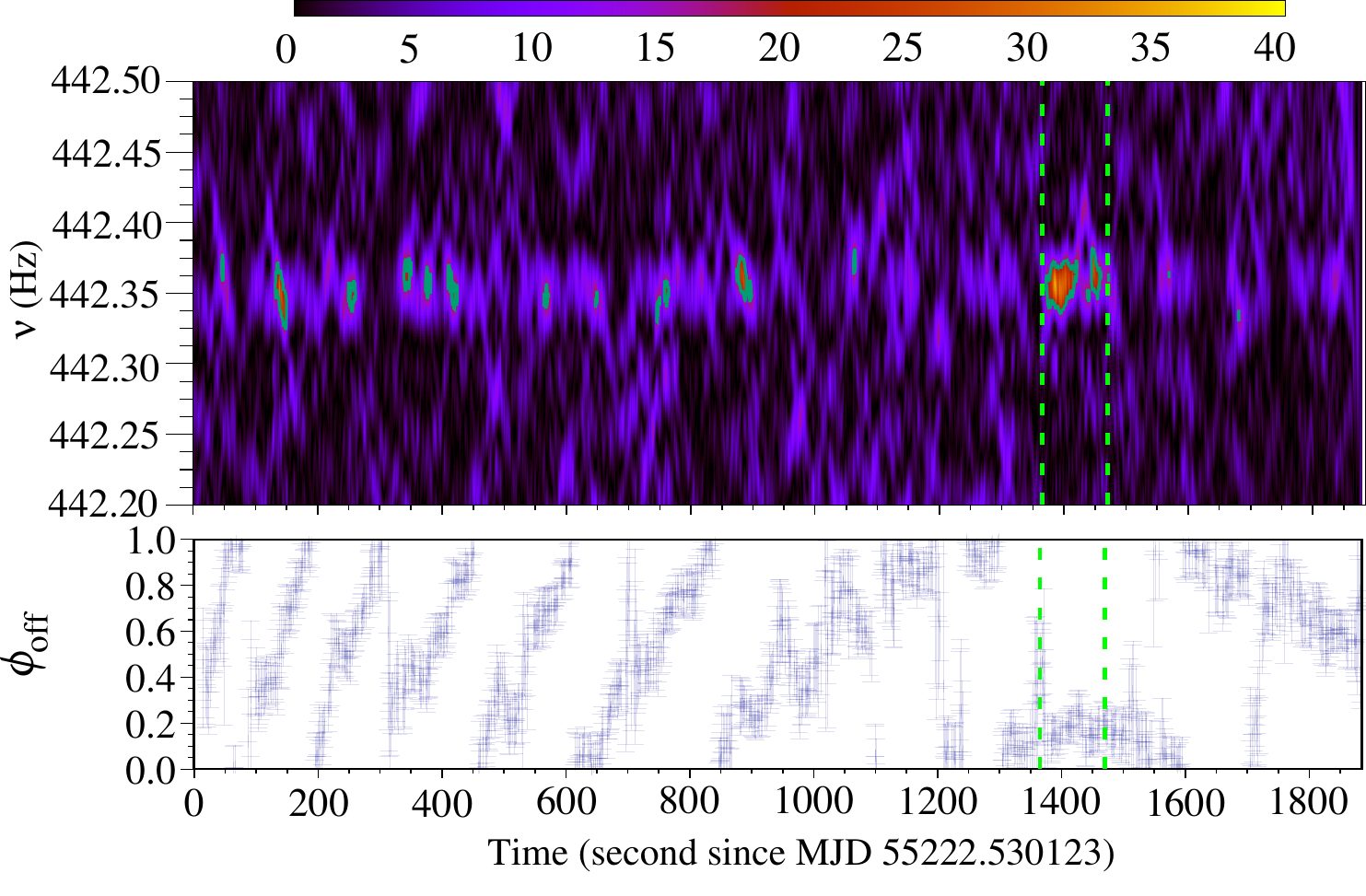}
  \includegraphics[width=3.90in, height=2.76in]{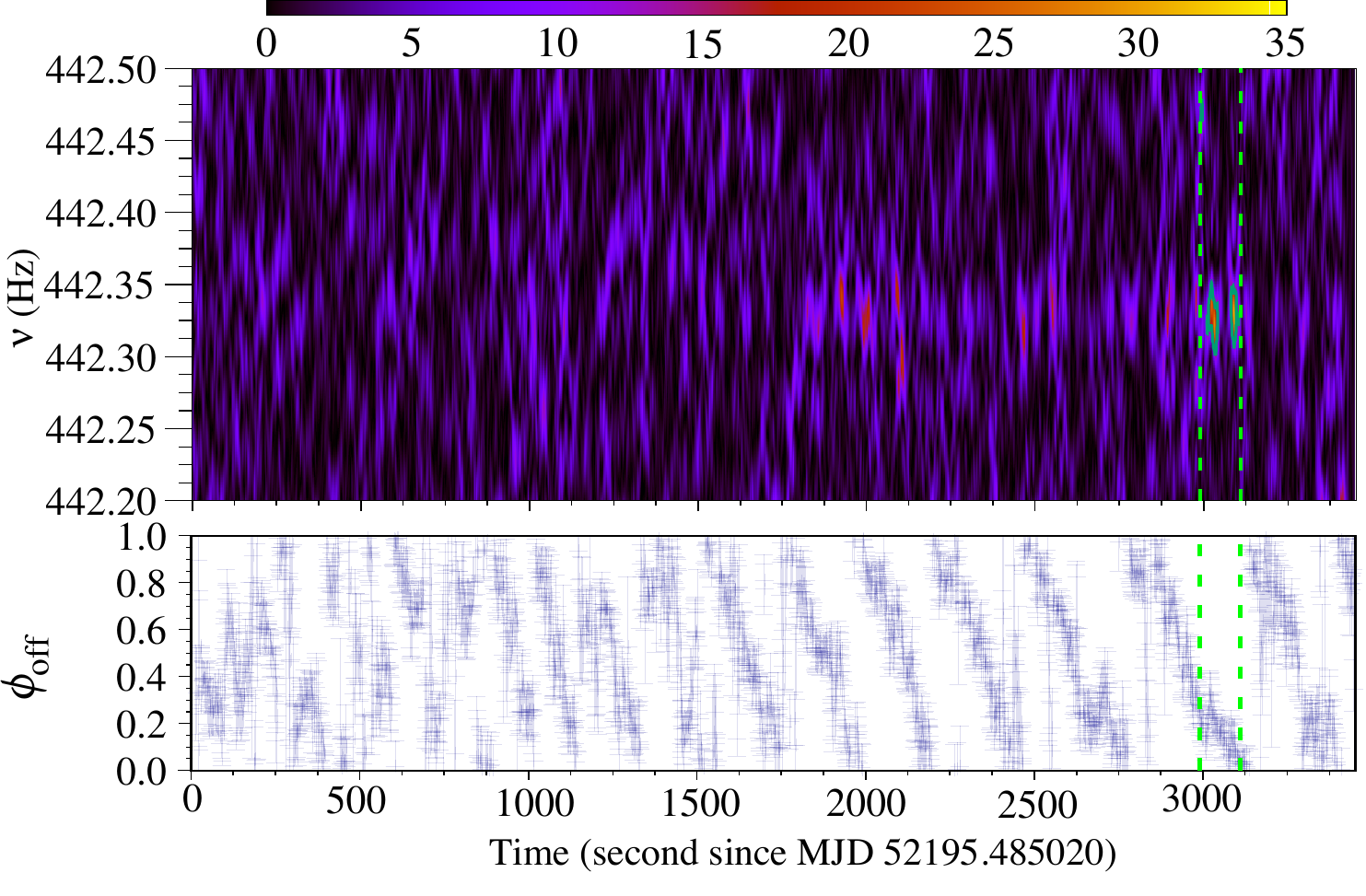}
  \caption{Same as \autoref{fig:aql-sax-literature}  but for two data sets of \sax; ObsID 94315-01-06-07 (upper) and ObsID: 60035-02-03-00 (bottom).}
\label{fig:aql-sax}
\end{figure}

We show dynamic power spectra and the time evolution of the phase shifts of two pulse candidates 
from \sax~(ObsID:94315-01-06-07, 60035-02-03-00) in \autoref{fig:aql-sax}. 
Unlikely the previous examples of the pulse candidates given in \autoref{pow-all-color} and \autoref{fig:aql-sax-literature},
the systematic changes in the pulse offset evolution, due to the orbital movement of the binary, clearly indicate that
the pulse durations are longer than the ones obtained via Z$_1^2$ method. This implies that the possible 
smoothness of the transitions between the pulse-on and pulse-off stages may show diversity.

The pulse phase shifts obtained from ML method are naturally between 0 and 1 in which the full cycle systematic 
shift makes the phase offsets return to 0.
One can simple modify $\phi_\text{off}$ values as each 0 to 1 order follows the next and is followed by the previous ones.
These systematic changes might be due to the orbital movement of the NS around the joint mass center of the binary system.
Therefore, the periodic variation represent the orbital period of the system.
On the other hand, if the orbital period is already known, the period obtained from the time evolution of the 
modified $\phi_\text{off}$ can be use to strengthen
the pulse detection via ML method even if Z$_1^2$ does not give any meaningful detection. 
On this basis, after we performed the Z$_1^2$ scan and obtained the pulse 
candidates, we performed the ML scan $\sim$50 days 
around them and tracked the possible systematic $\phi_\text{off}$ changes.

\newgeometry{top=0.8in} 

\begin{figure}[H]
\centering
  \includegraphics[width=6.5in, height=6in]{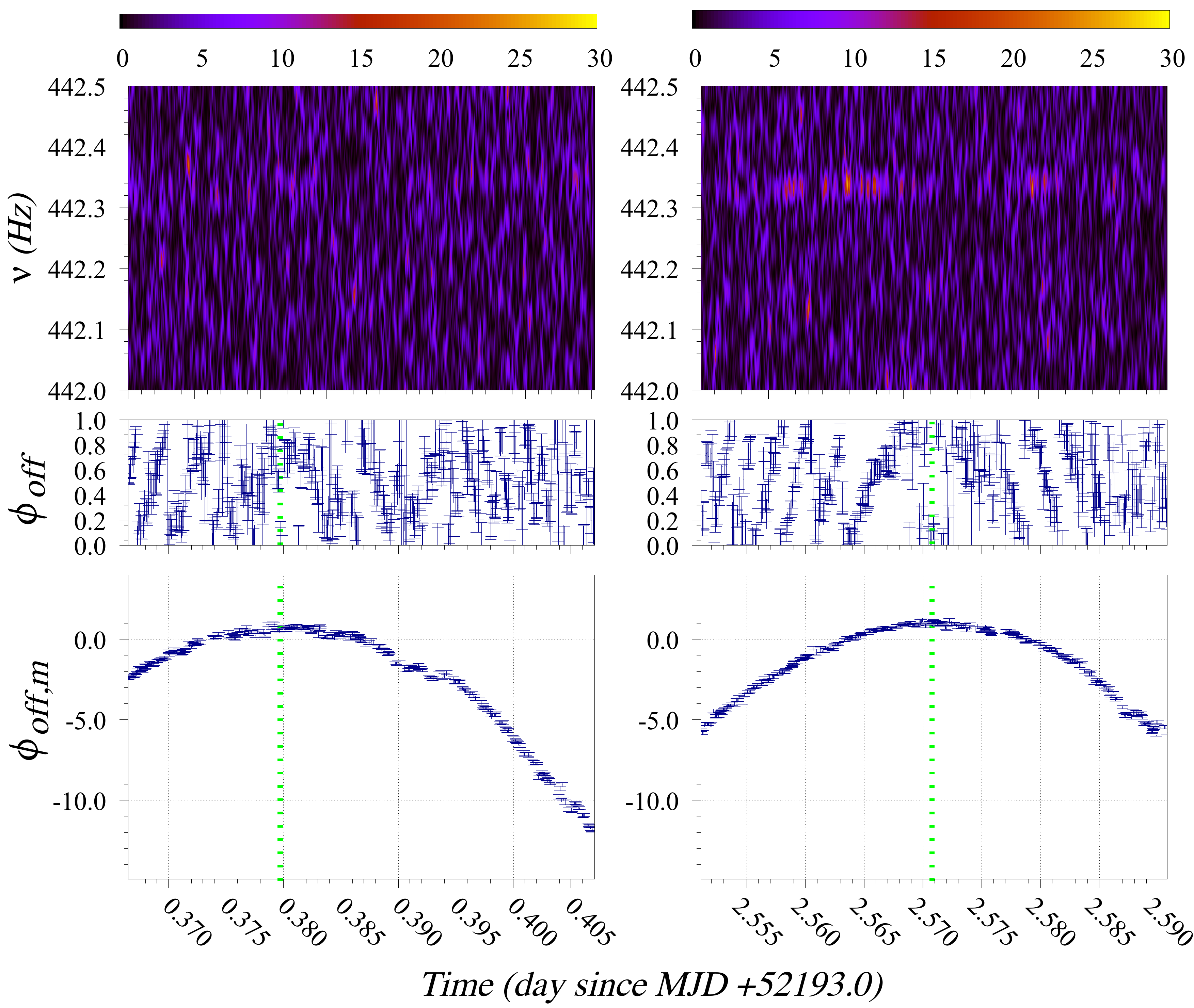}
\caption{The dynamic power spectra (upper panels), the time evolution of the phase shifts obtained via \ac{ML} (middle panels), and the time evolution of the modified phase offset values (bottom panels) for the two datasets: ObsID 60035-02-02-06 (left) and ObsID 6035-02-03-00 (right) of \sax. These are used to determine the orbital period of the system.}
\label{fig:sax-yorunge}
\end{figure}

We detected a pulse candidate, fit in our search criteria, in the data of ObsID 60035-02-03-00 
whose dynamic power spectrum is given in the upper right panel of \autoref{fig:sax-yorunge}. 
By using the pulse profile obtained from the corresponding time segment, the probability density function has been built
and pulse scan has been performed via ML method centering the time of the pulse as generally explained above.
The time evolution of the most probably phase offset values is shown in the middle panel. 

\newgeometry{left=1in, right=1in, top=0.8in, bottom=1in} 

We noticed a clear systematic change not only during the pulse duration given
by Z$_1^2$ but also covering the whole data as well as in the previous RXTE data, ObsId 60035-02-02-06.
The dynamic power spectrum of this data (upper left panel of \autoref{fig:sax-yorunge}) 
has no any clue of pulsation which is discussed in the upcoming paragraphs.
Phase offset values are, of course, between 0 and 1.
When we modify the phase offset values as each row in the systematical pattern follows the next one, 
we obtain two parabolas, that are a part of the expected sinusoidal shape due to the orbital movement of the system.
The turn over points seen in the bottom panels of \autoref{fig:sax-yorunge} are maxima of that sinus curve.
The turn points are determined as MJD 52193.379725 $\pm$ 0.000181 and  52195.570771 
$\pm$ 0.000096 for the ObsIDs 60035-02-02 06, 60035-02-03-00, respectively. Using the orbital period of the system
taken from the literature, we get how many cycles passed between these two maxima.
Diving the time gap with the cycle count we updated the orbital period of \sax~to be 8.764 ± 0.001 hours.

The \ac{ML} method had already been tested in other systems such as magnetars and remarked as more 
sensitive than the classical Fourier 
technique. We showed that ML method is also more precise than Z$_n^2$ test.
We used ML method to pulse scan of intermittent-\acp{AMXP} for the first time in our study.
The dynamic power spectra obtained by using the Z$^2$ method and the time variation of the phase offset values obtained by the M
method indicate 3 possible transition patterns between \textit{pulse-on} and \textit{pulse-off} stages:\\
\textit{(i)} The transition is sharp and show step function like pattern.
In \autoref{fig:aql-sax-literature}, the pulse transitions between the two stages present instant switches
as outcomes of both Z$_1^2$ and ML scan. In this case, the pulses are intermittent with almost constant amplitude.
One of the physical interpretations of such a transition may be related to the low accretion regime 
in the thin disc assumption when the inner radius of the disc is around the co-rotation radius.
Another discussion in the literature on such transitions relies on that
nearly aligned axes of the the magnetic field of the source and rotation can cause instantaneous transition
resulting pulsations to disappear for short periods \citep{Lamb2009a, Lamb2009b}. 
\citet{Kulkarni2008} argue that in some cases with high accretion, multiple plasma ``tongues'' can impact the magnetosphere and cause 
the X-ray signal to decay at indiscriminate locations on the NS surface.\\
\textit{(ii)} A clear smooth transition between \textit{pulse-on} and \textit{pulse-off} is seen in \autoref{fig:aql-sax}.
The pulses are again intermittent with a gentle switch as a result of the variation of pulse amplitude.
The pulse duration seen in the maximum likelihood output is clearly longer than the one given by $Z_1^2$.
The inhomogeneous comptonized corona around the neutron star can cause this soft transition.
The geometry and the homogeneity of the corona in these systems depend on the accretion rate, though,
the presence of pulsation is also expected to be correlated with the accretion rate.
On the other hand, \citet{Gogus2007} showed that the optical depth of the corona ($\tau$) is not thick enough to smear out the pulse.
Another explanation for the smooth transition could be the thick disc assumption.
While the inner radius of the disc plane is around the co-rotation radius, accretion can continue
from higher scale heights of the inner radius of the disc \citep{Gungor2017}.
\citep{Cumming2001} argue that high accretion rates lead to decay of the magnetic field which cause to weaken the pulse amplitude.
Small variations in the magnetic field lines shifting from the equilibrium position 
on the accretion disc can create turbulence \citep{Romanova2008} resulting small fluctuations in the accretion rate.\\
\textit{(iii)} The pulse might be continuous, but the amplitude of the pulse may vary over time to undetectable levels.
The possible scenarios behind this are the same as in the case of a soft transition.
A good example is the way how we obtained the orbital period of \sax.
The time evolution of the phase offset values throughout ML, consistent with the orbital period,
the pulse is clearly ongoing, while the dynamic power spectrum has no hint of any pulse (\autoref{fig:sax-yorunge}).

In conclusion, we cannot mention a universal transition between \textit{pulse-on} and \textit{pulse-off} stages. So that,
the smoothness of the transition is not unique for a system, but varies from event to event.

\section{Summary and Conclusions}
\label{summary}

We first perform a detailed pulse scan of the three intermittent-\acp{AMXP} using the Z$_1^2$ method.
We set a common threshold for Z$^{2}$ power values to define a detection as pulse candidates.
We obtained the precise values of T$_{0}$ and $\nu$ via the detection throughout Z$_1^2$ and used them to get the pulse profile to 
be used as a probability density function in the ML method. Following the ML procedure described above, the regions where the 
phase offsets show a systematic change in time are introduced as the new discovered pulse areas of the mentioned
intermittent-AMXPs.
The frequency and Z$^{2}$ power values of the pulse scan obtained for these three sources are given in \autoref{table1}.
The scans additionally allowed us to determine the pulse duration.

With results of our pulse scan, we can  comment on the intermittency and how the pulse phenomenon occurs.
From the figures discussed above, we can conclude that the pulse duration are longer than the ones previously presented in the 
literature. We show that the transition between the \textit{pulse-on} and \textit{pulse-off} stages are smooth and the smoothness varies.

Finally, we determined the orbital period of \sax~to be 8.764 $\pm$ 0.001 hours from the time evolution of the phase offsets
obtained via ML method in the time range when Z$_1^2$ method does not give any pulse detection. The methods that we used in this study might be applied to other data sets obtained from other X-ray missions.

\section*{Acknowledgements}
The authors would like to thank anonymous referee for her/his constructive comments \& suggestions.
We also would like to present great appretiation to Z. Funda Bostancı for usefull discussions.
This study is a part of the project titled ``Searching for Possible Pulsations in Low Mass X-ray Binaries via Maximum Likelihood 
Technique'' supported by TUBITAK (120F094). This work is supported by the Academy of Science Young Scientists Award Program (BAGEP). CG would like to thank the Academy of Science for their support.


\bibliographystyle{elsarticle-harv}

\begin{thebibliography}{33}
\expandafter\ifx\csname natexlab\endcsname\relax\def\natexlab#1{#1}\fi
\providecommand{\url}[1]{\texttt{#1}}
\providecommand{\href}[2]{#2}
\providecommand{\path}[1]{#1}
\providecommand{\DOIprefix}{doi:}
\providecommand{\ArXivprefix}{arXiv:}
\providecommand{\URLprefix}{URL: }
\providecommand{\Pubmedprefix}{pmid:}
\providecommand{\doi}[1]{\href{http://dx.doi.org/#1}{\path{#1}}}
\providecommand{\Pubmed}[1]{\href{pmid:#1}{\path{#1}}}
\providecommand{\bibinfo}[2]{#2}
\ifx\xfnm\relax \def\xfnm[#1]{\unskip,\space#1}\fi
\bibitem[{{Alpar} et~al.(1982){Alpar}, {Cheng}, {Ruderman} and
  {Shaham}}]{1982Natur.300..728A}
\bibinfo{author}{{Alpar}, M.A.}, \bibinfo{author}{{Cheng}, A.F.},
  \bibinfo{author}{{Ruderman}, M.A.}, \bibinfo{author}{{Shaham}, J.},
  \bibinfo{year}{1982}.
\newblock \bibinfo{title}{{A new class of radio pulsars}}.
\newblock \bibinfo{journal}{\nat} \bibinfo{volume}{300},
  \bibinfo{pages}{728--730}.
\newblock \DOIprefix\doi{10.1038/300728a0}.
\bibitem[{{Alpar} and {Shaham}(1985)}]{1985Natur.316..239A}
\bibinfo{author}{{Alpar}, M.A.}, \bibinfo{author}{{Shaham}, J.},
  \bibinfo{year}{1985}.
\newblock \bibinfo{title}{{Is GX5 - 1 a millisecond pulsar?}}
\newblock \bibinfo{journal}{\nat} \bibinfo{volume}{316},
  \bibinfo{pages}{239--241}.
\newblock \DOIprefix\doi{10.1038/316239a0}.
\bibitem[{{Altamirano} et~al.(2008){Altamirano}, {Casella}, {Patruno},
  {Wijnands} and {van der Klis}}]{AltamiranoSax2008}
\bibinfo{author}{{Altamirano}, D.}, \bibinfo{author}{{Casella}, P.},
  \bibinfo{author}{{Patruno}, A.}, \bibinfo{author}{{Wijnands}, R.},
  \bibinfo{author}{{van der Klis}, M.}, \bibinfo{year}{2008}.
\newblock \bibinfo{title}{{Intermittent Millisecond X-Ray Pulsations from the
  Neutron Star X-Ray Transient SAX J1748.9-2021 in the Globular Cluster NGC
  6440}}.
\newblock \bibinfo{journal}{\apjl} \bibinfo{volume}{674}, \bibinfo{pages}{L45}.
\newblock \DOIprefix\doi{10.1086/528983},
  \href{http://arxiv.org/abs/0708.1316}{{\tt arXiv:0708.1316}}.
\bibitem[{{Bahar} et~al.(2021){Bahar}, {Chakraborty} and
  {G{\"o}{\v{g}}{\"u}{\c{s}}}}]{2021PASA...38...11B}
\bibinfo{author}{{Bahar}, Y.E.}, \bibinfo{author}{{Chakraborty}, M.},
  \bibinfo{author}{{G{\"o}{\v{g}}{\"u}{\c{s}}}, E.}, \bibinfo{year}{2021}.
\newblock \bibinfo{title}{{Search for intermittent X-ray pulsations from
  neutron stars in low-mass X-ray binaries}}.
\newblock \bibinfo{journal}{\pasa} \bibinfo{volume}{38}, \bibinfo{pages}{e011}.
\newblock \DOIprefix\doi{10.1017/pasa.2021.6},
  \href{http://arxiv.org/abs/2102.03132}{{\tt arXiv:2102.03132}}.
\bibitem[{{Brainerd} and {Lamb}(1987)}]{1987ApJ...317L..33B}
\bibinfo{author}{{Brainerd}, J.}, \bibinfo{author}{{Lamb}, F.K.},
  \bibinfo{year}{1987}.
\newblock \bibinfo{title}{{Effect of an Electron Scattering Cloud on X-Ray
  Oscillations Produced by Beaming}}.
\newblock \bibinfo{journal}{\apjl} \bibinfo{volume}{317}, \bibinfo{pages}{L33}.
\newblock \DOIprefix\doi{10.1086/184908}.
\bibitem[{{Buccheri} et~al.(1983){Buccheri}, {Bennett}, {Bignami}, {Bloemen},
  {Boriakoff}, {Caraveo}, {Hermsen}, {Kanbach}, {Manchester}, {Masnou},
  {Mayer-Hasselwander}, {{\"O}zel}, {Paul}, {Sacco}, {Scarsi} and
  {Strong}}]{1983A&A...128..245B}
\bibinfo{author}{{Buccheri}, R.}, \bibinfo{author}{{Bennett}, K.},
  \bibinfo{author}{{Bignami}, G.F.}, \bibinfo{author}{{Bloemen}, J.B.G.M.},
  \bibinfo{author}{{Boriakoff}, V.}, \bibinfo{author}{{Caraveo}, P.A.},
  \bibinfo{author}{{Hermsen}, W.}, \bibinfo{author}{{Kanbach}, G.},
  \bibinfo{author}{{Manchester}, R.N.}, \bibinfo{author}{{Masnou}, J.L.},
  \bibinfo{author}{{Mayer-Hasselwander}, H.A.}, \bibinfo{author}{{{\"O}zel},
  M.E.}, \bibinfo{author}{{Paul}, J.A.}, \bibinfo{author}{{Sacco}, B.},
  \bibinfo{author}{{Scarsi}, L.}, \bibinfo{author}{{Strong}, A.W.},
  \bibinfo{year}{1983}.
\newblock \bibinfo{title}{{Search for pulsed {\ensuremath{\gamma}}-ray emission
  from radio pulsars in the COS-B data.}}
\newblock \bibinfo{journal}{\aap} \bibinfo{volume}{128},
  \bibinfo{pages}{245--251}.
\bibitem[{{Burderi} and {Di Salvo}(2013)}]{2013MmSAI..84..117B}
\bibinfo{author}{{Burderi}, L.}, \bibinfo{author}{{Di Salvo}, T.},
  \bibinfo{year}{2013}.
\newblock \bibinfo{title}{{On low mass X-ray binaries and millisecond pulsar}}.
\newblock \bibinfo{journal}{\memsai} \bibinfo{volume}{84},
  \bibinfo{pages}{117}.
\newblock \DOIprefix\doi{10.48550/arXiv.1310.1283},
  \href{http://arxiv.org/abs/1310.1283}{{\tt arXiv:1310.1283}}.
\bibitem[{{Casella} et~al.(2008){Casella}, {Altamirano}, {Patruno}, {Wijnands}
  and {van der Klis}}]{Casella2008}
\bibinfo{author}{{Casella}, P.}, \bibinfo{author}{{Altamirano}, D.},
  \bibinfo{author}{{Patruno}, A.}, \bibinfo{author}{{Wijnands}, R.},
  \bibinfo{author}{{van der Klis}, M.}, \bibinfo{year}{2008}.
\newblock \bibinfo{title}{{Discovery of Coherent Millisecond X-Ray Pulsations
  in Aquila X-1}}.
\newblock \bibinfo{journal}{\apjl} \bibinfo{volume}{674}, \bibinfo{pages}{L41}.
\newblock \DOIprefix\doi{10.1086/528982},
  \href{http://arxiv.org/abs/0708.1110}{{\tt arXiv:0708.1110}}.
\bibitem[{{Chaty}(2022)}]{Chaty2022}
\bibinfo{author}{{Chaty}, S.}, \bibinfo{year}{2022}.
\newblock \bibinfo{title}{{Accreting Binaries; Nature, formation, and
  evolution}}.
\newblock \DOIprefix\doi{10.1088/2514-3433/ac595f}.
\bibitem[{{Chevalier} and {Ilovaisky}(1991)}]{1991A&A...251L..11C}
\bibinfo{author}{{Chevalier}, C.}, \bibinfo{author}{{Ilovaisky}, S.A.},
  \bibinfo{year}{1991}.
\newblock \bibinfo{title}{{Discovery of a 19-hour period in Aquila X-1.}}
\newblock \bibinfo{journal}{\aap} \bibinfo{volume}{251}, \bibinfo{pages}{L11}.
\bibitem[{{Cumming} et~al.(2001){Cumming}, {Zweibel} and
  {Bildsten}}]{Cumming2001}
\bibinfo{author}{{Cumming}, A.}, \bibinfo{author}{{Zweibel}, E.},
  \bibinfo{author}{{Bildsten}, L.}, \bibinfo{year}{2001}.
\newblock \bibinfo{title}{{Magnetic Screening in Accreting Neutron Stars}}.
\newblock \bibinfo{journal}{\apj} \bibinfo{volume}{557},
  \bibinfo{pages}{958--966}.
\newblock \DOIprefix\doi{10.1086/321658},
  \href{http://arxiv.org/abs/astro-ph/0102178}{{\tt arXiv:astro-ph/0102178}}.
\bibitem[{{Fortin} et~al.(2024){Fortin}, {Kalsi}, {Garc{\'\i}a} and
  {Chaty}}]{Fortin2024}
\bibinfo{author}{{Fortin}, F.}, \bibinfo{author}{{Kalsi}, A.},
  \bibinfo{author}{{Garc{\'\i}a}, F.}, \bibinfo{author}{{Chaty}, S.},
  \bibinfo{year}{2024}.
\newblock \bibinfo{title}{{A catalogue of low-mass X-ray binaries in the
  Galaxy: from the INTEGRAL to the Gaia era}}.
\newblock \bibinfo{journal}{arXiv e-prints} ,
  \bibinfo{pages}{arXiv:2401.11931}\href{http://arxiv.org/abs/2401.11931}{{\tt
  arXiv:2401.11931}}.
\bibitem[{{Galloway} et~al.(2008){Galloway}, {Muno}, {Hartman}, {Psaltis} and
  {Chakrabarty}}]{GallowayThermo2008}
\bibinfo{author}{{Galloway}, D.K.}, \bibinfo{author}{{Muno}, M.P.},
  \bibinfo{author}{{Hartman}, J.M.}, \bibinfo{author}{{Psaltis}, D.},
  \bibinfo{author}{{Chakrabarty}, D.}, \bibinfo{year}{2008}.
\newblock \bibinfo{title}{{Thermonuclear (Type I) X-Ray Bursts Observed by the
  Rossi X-Ray Timing Explorer}}.
\newblock \bibinfo{journal}{\apjs} \bibinfo{volume}{179},
  \bibinfo{pages}{360--422}.
\newblock \DOIprefix\doi{10.1086/592044},
  \href{http://arxiv.org/abs/astro-ph/0608259}{{\tt arXiv:astro-ph/0608259}}.
\bibitem[{{Gavriil} et~al.(2007){Gavriil}, {Strohmayer}, {Swank} and
  {Markwardt}}]{2007ApJ...669L..29G}
\bibinfo{author}{{Gavriil}, F.P.}, \bibinfo{author}{{Strohmayer}, T.E.},
  \bibinfo{author}{{Swank}, J.H.}, \bibinfo{author}{{Markwardt}, C.B.},
  \bibinfo{year}{2007}.
\newblock \bibinfo{title}{{Discovery of 442 Hz Pulsations from an X-Ray Source
  in the Globular Cluster NGC 6440}}.
\newblock \bibinfo{journal}{\apjl} \bibinfo{volume}{669},
  \bibinfo{pages}{L29--L32}.
\newblock \DOIprefix\doi{10.1086/523758},
  \href{http://arxiv.org/abs/0708.0829}{{\tt arXiv:0708.0829}}.
\bibitem[{{G{\"o}{\v{g}}{\"u}{\c{s}}} et~al.(2007){G{\"o}{\v{g}}{\"u}{\c{s}}},
  {Alpar} and {Gilfanov}}]{Gogus2007}
\bibinfo{author}{{G{\"o}{\v{g}}{\"u}{\c{s}}}, E.}, \bibinfo{author}{{Alpar},
  M.A.}, \bibinfo{author}{{Gilfanov}, M.}, \bibinfo{year}{2007}.
\newblock \bibinfo{title}{{Is the Lack of Pulsations in Low-Mass X-Ray Binaries
  due to Comptonizing Coronae?}}
\newblock \bibinfo{journal}{\apj} \bibinfo{volume}{659},
  \bibinfo{pages}{580--584}.
\newblock \DOIprefix\doi{10.1086/512028},
  \href{http://arxiv.org/abs/astro-ph/0612680}{{\tt arXiv:astro-ph/0612680}}.
\bibitem[{{G{\"u}ng{\"o}r} et~al.(2017){G{\"u}ng{\"o}r}, {Ek{\c{s}}i},
  {G{\"o}{\u{g}}{\"u}{\c{s}}} and {G{\"u}ver}}]{Gungor2017}
\bibinfo{author}{{G{\"u}ng{\"o}r}, C.}, \bibinfo{author}{{Ek{\c{s}}i}, K.Y.},
  \bibinfo{author}{{G{\"o}{\u{g}}{\"u}{\c{s}}}, E.},
  \bibinfo{author}{{G{\"u}ver}, T.}, \bibinfo{year}{2017}.
\newblock \bibinfo{title}{{Partial Accretion in the Propeller Stage of Low-mass
  X-Ray Binary Aql X-1}}.
\newblock \bibinfo{journal}{\apj} \bibinfo{volume}{848}, \bibinfo{pages}{13}.
\newblock \DOIprefix\doi{10.3847/1538-4357/aa8b76},
  \href{http://arxiv.org/abs/1709.02378}{{\tt arXiv:1709.02378}}.
\bibitem[{{in 't Zand} et~al.(1999){in 't Zand}, {Verbunt}, {Strohmayer},
  {Bazzano}, {Cocchi}, {Heise}, {van Kerkwijk}, {Muller}, {Natalucci}, {Smith}
  and {Ubertini}}]{1999A&A...345..100I}
\bibinfo{author}{{in 't Zand}, J.J.M.}, \bibinfo{author}{{Verbunt}, F.},
  \bibinfo{author}{{Strohmayer}, T.E.}, \bibinfo{author}{{Bazzano}, A.},
  \bibinfo{author}{{Cocchi}, M.}, \bibinfo{author}{{Heise}, J.},
  \bibinfo{author}{{van Kerkwijk}, M.H.}, \bibinfo{author}{{Muller}, J.M.},
  \bibinfo{author}{{Natalucci}, L.}, \bibinfo{author}{{Smith}, M.J.S.},
  \bibinfo{author}{{Ubertini}, P.}, \bibinfo{year}{1999}.
\newblock \bibinfo{title}{{A new X-ray outburst in the globular cluster NGC
  6440: SAX J1748.9-2021}}.
\newblock \bibinfo{journal}{\aap} \bibinfo{volume}{345},
  \bibinfo{pages}{100--108}.
\newblock \DOIprefix\doi{10.48550/arXiv.astro-ph/9902319},
  \href{http://arxiv.org/abs/astro-ph/9902319}{{\tt arXiv:astro-ph/9902319}}.
\bibitem[{Jonker and Nelemans(2004)}]{jonker2004distances}
\bibinfo{author}{Jonker, P.G.}, \bibinfo{author}{Nelemans, G.},
  \bibinfo{year}{2004}.
\newblock \bibinfo{title}{The distances to galactic low-mass x-ray binaries:
  consequences for black hole luminosities and kicks}.
\newblock \bibinfo{journal}{Monthly Notices of the Royal Astronomical Society}
  \bibinfo{volume}{354}, \bibinfo{pages}{355--366}.
\bibitem[{{Kaaret} et~al.(2006){Kaaret}, {Morgan}, {Vanderspek} and
  {Tomsick}}]{2006ApJ...638..963K}
\bibinfo{author}{{Kaaret}, P.}, \bibinfo{author}{{Morgan}, E.H.},
  \bibinfo{author}{{Vanderspek}, R.}, \bibinfo{author}{{Tomsick}, J.A.},
  \bibinfo{year}{2006}.
\newblock \bibinfo{title}{{Discovery of the Millisecond X-Ray Pulsar HETE
  J1900.1-2455}}.
\newblock \bibinfo{journal}{\apj} \bibinfo{volume}{638},
  \bibinfo{pages}{963--967}.
\newblock \DOIprefix\doi{10.1086/498886},
  \href{http://arxiv.org/abs/astro-ph/0510483}{{\tt arXiv:astro-ph/0510483}}.
\bibitem[{Koyama et~al.(1981)Koyama, Inoue, Makishima, Matsuoka, Murakami, Oda,
  Osgawara, Ohashi, Shibazaki, Tanaka et~al.}]{koyama1981discovery}
\bibinfo{author}{Koyama, K.}, \bibinfo{author}{Inoue, H.},
  \bibinfo{author}{Makishima, K.}, \bibinfo{author}{Matsuoka, M.},
  \bibinfo{author}{Murakami, T.}, \bibinfo{author}{Oda, M.},
  \bibinfo{author}{Osgawara, Y.}, \bibinfo{author}{Ohashi, T.},
  \bibinfo{author}{Shibazaki, N.}, \bibinfo{author}{Tanaka, Y.}, et~al.,
  \bibinfo{year}{1981}.
\newblock \bibinfo{title}{Discovery of x-ray bursts from aquila x-1}.
\newblock \bibinfo{journal}{The Astrophysical Journal} \bibinfo{volume}{247},
  \bibinfo{pages}{L27--L29}.
\bibitem[{{Kulkarni} and {Romanova}(2008)}]{Kulkarni2008}
\bibinfo{author}{{Kulkarni}, A.K.}, \bibinfo{author}{{Romanova}, M.M.},
  \bibinfo{year}{2008}.
\newblock \bibinfo{title}{{Accretion to magnetized stars through the
  Rayleigh-Taylor instability: global 3D simulations}}.
\newblock \bibinfo{journal}{\mnras} \bibinfo{volume}{386},
  \bibinfo{pages}{673--687}.
\newblock \DOIprefix\doi{10.1111/j.1365-2966.2008.13094.x},
  \href{http://arxiv.org/abs/0802.1759}{{\tt arXiv:0802.1759}}.
\bibitem[{{Kylafis}(1988)}]{1988AdSpR...8b.455K}
\bibinfo{author}{{Kylafis}, N.D.}, \bibinfo{year}{1988}.
\newblock \bibinfo{title}{{Temporal effects of electron scattering on the
  oscillations of an X-ray source}}.
\newblock \bibinfo{journal}{Advances in Space Research} \bibinfo{volume}{8},
  \bibinfo{pages}{455--458}.
\newblock \DOIprefix\doi{10.1016/0273-1177(88)90442-5}.
\bibitem[{{Lamb} et~al.(2009a){Lamb}, {Boutloukos}, {Van Wassenhove},
  {Chamberlain}, {Lo}, {Clare}, {Yu} and {Miller}}]{Lamb2009b}
\bibinfo{author}{{Lamb}, F.K.}, \bibinfo{author}{{Boutloukos}, S.},
  \bibinfo{author}{{Van Wassenhove}, S.}, \bibinfo{author}{{Chamberlain},
  R.T.}, \bibinfo{author}{{Lo}, K.H.}, \bibinfo{author}{{Clare}, A.},
  \bibinfo{author}{{Yu}, W.}, \bibinfo{author}{{Miller}, M.C.},
  \bibinfo{year}{2009}a.
\newblock \bibinfo{title}{{A Model for the Waveform Behavior of Accreting
  Millisecond X-Ray Pulsars: Nearly Aligned Magnetic Fields and Moving Emission
  Regions}}.
\newblock \bibinfo{journal}{\apj} \bibinfo{volume}{706},
  \bibinfo{pages}{417--435}.
\newblock \DOIprefix\doi{10.1088/0004-637X/706/1/417},
  \href{http://arxiv.org/abs/0808.4159}{{\tt arXiv:0808.4159}}.
\bibitem[{{Lamb} et~al.(2009b){Lamb}, {Boutloukos}, {Van Wassenhove},
  {Chamberlain}, {Lo} and {Miller}}]{Lamb2009a}
\bibinfo{author}{{Lamb}, F.K.}, \bibinfo{author}{{Boutloukos}, S.},
  \bibinfo{author}{{Van Wassenhove}, S.}, \bibinfo{author}{{Chamberlain},
  R.T.}, \bibinfo{author}{{Lo}, K.H.}, \bibinfo{author}{{Miller}, M.C.},
  \bibinfo{year}{2009}b.
\newblock \bibinfo{title}{{Origin of Intermittent Accretion-Powered X-ray
  Oscillations in Neutron Stars with Millisecond Spin Periods}}.
\newblock \bibinfo{journal}{\apjl} \bibinfo{volume}{705},
  \bibinfo{pages}{L36--L39}.
\newblock \DOIprefix\doi{10.1088/0004-637X/705/1/L36},
  \href{http://arxiv.org/abs/0809.4016}{{\tt arXiv:0809.4016}}.
\bibitem[{{Livingstone} et~al.(2009){Livingstone}, {Ransom}, {Camilo}, {Kaspi},
  {Lyne}, {Kramer} and {Stairs}}]{2009ApJ...706.1163L}
\bibinfo{author}{{Livingstone}, M.A.}, \bibinfo{author}{{Ransom}, S.M.},
  \bibinfo{author}{{Camilo}, F.}, \bibinfo{author}{{Kaspi}, V.M.},
  \bibinfo{author}{{Lyne}, A.G.}, \bibinfo{author}{{Kramer}, M.},
  \bibinfo{author}{{Stairs}, I.H.}, \bibinfo{year}{2009}.
\newblock \bibinfo{title}{{X-ray and Radio Timing of the Pulsar in 3C 58}}.
\newblock \bibinfo{journal}{\apj} \bibinfo{volume}{706},
  \bibinfo{pages}{1163--1173}.
\newblock \DOIprefix\doi{10.1088/0004-637X/706/2/1163},
  \href{http://arxiv.org/abs/0901.2119}{{\tt arXiv:0901.2119}}.
\bibitem[{Mata~Sánchez et~al.(2016)Mata~Sánchez, Muñoz-Darias, Casares and
  Jiménez-Ibarra}]{10.1093/mnrasl/slw172}
\bibinfo{author}{Mata~Sánchez, D.}, \bibinfo{author}{Muñoz-Darias, T.},
  \bibinfo{author}{Casares, J.}, \bibinfo{author}{Jiménez-Ibarra, F.},
  \bibinfo{year}{2016}.
\newblock \bibinfo{title}{{The donor of Aquila X-1 revealed by high-angular
  resolution near-infrared spectroscopy}}.
\newblock \bibinfo{journal}{Monthly Notices of the Royal Astronomical Society:
  Letters} \bibinfo{volume}{464}, \bibinfo{pages}{L41--L45}.
\newblock \URLprefix \url{https://doi.org/10.1093/mnrasl/slw172},
  \DOIprefix\doi{10.1093/mnrasl/slw172}.
\bibitem[{{Meszaros} and {Riffert}(1988)}]{1988ApJ...327..712M}
\bibinfo{author}{{Meszaros}, P.}, \bibinfo{author}{{Riffert}, H.},
  \bibinfo{year}{1988}.
\newblock \bibinfo{title}{{Gravitational Light Bending near Neutron Stars. II.
  Accreting Pulsar Spectra as a Function of Phase}}.
\newblock \bibinfo{journal}{\apj} \bibinfo{volume}{327}, \bibinfo{pages}{712}.
\newblock \DOIprefix\doi{10.1086/166227}.
\bibitem[{{Ortolani} et~al.(1994){Ortolani}, {Barbuy} and
  {Bica}}]{1994A&AS..108..653O}
\bibinfo{author}{{Ortolani}, S.}, \bibinfo{author}{{Barbuy}, B.},
  \bibinfo{author}{{Bica}, E.}, \bibinfo{year}{1994}.
\newblock \bibinfo{title}{{The low galactic latitude metal-rich globular
  cluster NGC 6440.}}
\newblock \bibinfo{journal}{\aaps} \bibinfo{volume}{108},
  \bibinfo{pages}{653--659}.
\bibitem[{{Patruno} et~al.(2009){Patruno}, {Altamirano}, {Hessels}, {Casella},
  {Wijnands} and {van der Klis}}]{2009ApJ...690.1856P}
\bibinfo{author}{{Patruno}, A.}, \bibinfo{author}{{Altamirano}, D.},
  \bibinfo{author}{{Hessels}, J.W.T.}, \bibinfo{author}{{Casella}, P.},
  \bibinfo{author}{{Wijnands}, R.}, \bibinfo{author}{{van der Klis}, M.},
  \bibinfo{year}{2009}.
\newblock \bibinfo{title}{{Phase-Coherent Timing of the Accreting Millisecond
  Pulsar SAX J1748.9-2021}}.
\newblock \bibinfo{journal}{\apj} \bibinfo{volume}{690},
  \bibinfo{pages}{1856--1865}.
\newblock \DOIprefix\doi{10.1088/0004-637X/690/2/1856},
  \href{http://arxiv.org/abs/0801.1031}{{\tt arXiv:0801.1031}}.
\bibitem[{{Romanova} et~al.(2008){Romanova}, {Kulkarni} and
  {Lovelace}}]{Romanova2008}
\bibinfo{author}{{Romanova}, M.M.}, \bibinfo{author}{{Kulkarni}, A.K.},
  \bibinfo{author}{{Lovelace}, R.V.E.}, \bibinfo{year}{2008}.
\newblock \bibinfo{title}{{Unstable Disk Accretion onto Magnetized Stars: First
  Global Three-dimensional Magnetohydrodynamic Simulations}}.
\newblock \bibinfo{journal}{\apjl} \bibinfo{volume}{673},
  \bibinfo{pages}{L171}.
\newblock \DOIprefix\doi{10.1086/527298},
  \href{http://arxiv.org/abs/0711.0418}{{\tt arXiv:0711.0418}}.
\bibitem[{{Sanna} et~al.(2016){Sanna}, {Burderi}, {Riggio}, {Pintore}, {Di
  Salvo}, {Gambino}, {Iaria}, {Matranga} and {Scarano}}]{2016MNRAS.459.1340S}
\bibinfo{author}{{Sanna}, A.}, \bibinfo{author}{{Burderi}, L.},
  \bibinfo{author}{{Riggio}, A.}, \bibinfo{author}{{Pintore}, F.},
  \bibinfo{author}{{Di Salvo}, T.}, \bibinfo{author}{{Gambino}, A.F.},
  \bibinfo{author}{{Iaria}, R.}, \bibinfo{author}{{Matranga}, M.},
  \bibinfo{author}{{Scarano}, F.}, \bibinfo{year}{2016}.
\newblock \bibinfo{title}{{Timing of the accreting millisecond pulsar SAX
  J1748.9-2021 during its 2015 outburst}}.
\newblock \bibinfo{journal}{\mnras} \bibinfo{volume}{459},
  \bibinfo{pages}{1340--1349}.
\newblock \DOIprefix\doi{10.1093/mnras/stw740},
  \href{http://arxiv.org/abs/1603.08757}{{\tt arXiv:1603.08757}}.
\bibitem[{{Scholz} et~al.(2014){Scholz}, {Archibald}, {Kaspi}, {Ng},
  {Beardmore}, {Gehrels} and {Kennea}}]{2014ApJ...783...99S}
\bibinfo{author}{{Scholz}, P.}, \bibinfo{author}{{Archibald}, R.F.},
  \bibinfo{author}{{Kaspi}, V.M.}, \bibinfo{author}{{Ng}, C.Y.},
  \bibinfo{author}{{Beardmore}, A.P.}, \bibinfo{author}{{Gehrels}, N.},
  \bibinfo{author}{{Kennea}, J.A.}, \bibinfo{year}{2014}.
\newblock \bibinfo{title}{{On the X-Ray Variability of Magnetar 1RXS
  J170849.0-400910}}.
\newblock \bibinfo{journal}{\apj} \bibinfo{volume}{783}, \bibinfo{pages}{99}.
\newblock \DOIprefix\doi{10.1088/0004-637X/783/2/99},
  \href{http://arxiv.org/abs/1401.5000}{{\tt arXiv:1401.5000}}.
\bibitem[{{Vanderspek} et~al.(2005){Vanderspek}, {Morgan}, {Crew}, {Graziani}
  and {Suzuki}}]{2005ATel..516....1V}
\bibinfo{author}{{Vanderspek}, R.}, \bibinfo{author}{{Morgan}, E.},
  \bibinfo{author}{{Crew}, G.}, \bibinfo{author}{{Graziani}, C.},
  \bibinfo{author}{{Suzuki}, M.}, \bibinfo{year}{2005}.
\newblock \bibinfo{title}{{Possible new X-ray burst source detected by HETE}}.
\newblock \bibinfo{journal}{The Astronomer's Telegram} \bibinfo{volume}{516},
  \bibinfo{pages}{1}.

\end{thebibliography}


\end{document}